\documentclass[12pt]{JHEP3}
\usepackage{amssymb,amsfonts}

\usepackage{amsmath}
\usepackage{graphicx}
\usepackage{amsmath,epsfig}
\usepackage{amssymb,amsfonts}
\usepackage{latexsym,bm}

\def\hri#1#2{\href{http://arxiv.org/abs/#1}{[ArXiv:#1]#2}}
\def\hre#1#2{\href{http://arxiv.org/abs/#1/#2}{[ArXiv:#1/#2]}}

\def\sp{\;\;\;,\;\;\;}

\newcommand{\be}{\begin{equation}}
\newcommand{\ee}{\end{equation}}
\newcommand{\bea}{\begin{eqnarray}}
\newcommand{\eea}{\end{eqnarray}}
\newcommand{\bem}{\begin{multline}}
\newcommand{\eem}{\end{multline}}
\newcommand{\beg}{\begin{gather}}
\newcommand{\eeg}{\end{gather}}

\newcommand{\ben}{\begin{eqnarray*}}
\newcommand{\een}{\end{eqnarray*}}

\title{\vskip -1cm Multiplicities from black-hole formation in heavy-ion collisions}
\author{Elias Kiritsis$^{1,2}$,  Anastasios Taliotis$^{1}$\\
~\\
 $^1$\href{http://hep.physics.uoc.gr/}{Crete Center for Theoretical Physics,\\Department of Physics, University of Crete
71003 Heraklion, Greece}\\
~\\
$^2$\href{http://www.apc.univ-paris7.fr}{APC, Universit\'e Paris 7, B\^atiment Condorcet, F-75205, Paris Cedex 13, France,
 (UMR du CNRS 7164).}
 }
\preprint{CCTP-2011-33}
\abstract{The formation of trapped surfaces in the head-on collision of shock waves in conformal and non-conformal backgrounds  is investigated. The backgrounds include all interesting confining and non-confining backgrounds that may be relevant for QCD.  Several transverse profiles of the shocks are investigated including distributions that fall-off as powers or exponentials.
Different ways of cutting-off the UV contributions (that are expected to be  perturbative in QCD) are explored.
Under some plausible simplifying assumptions our estimates are converted into predictions for multiplicities for heavy-ion collisions at RHIC and LHC.
}
\vspace{-1cm}

\begin{document}

%%%%%%%%%%%%%%%%%%%%%%%%%%%%%%%%%%%%%%%%%%%%%%%%%%%%%%%%%%%%%%%%%%%%%%%%%%%%%%%%%

\section{Introduction}

Data from heavy-ion collisions from RHIC and LHC have revolutionized our perception of strong coupling physics in QCD, and revealed the characteristics of the deconfined phase. They have also become the testing ground of novel techniques emerging in string theory that attempt to control strong coupling phenomena using a gravitational description.
In this description, a heavy-ion collision is described as the process of black hole formation and decay, albeit in a five-dimensional theory of gravity including also other fields, notable a scalar, the dilaton, \cite{kt}.

In this direction, several attempts have been made to analyse the scattering of high-energy sources, using shock waves in AdS, \cite{Kang:2004jd,Giddings:2002cd,Lin:2010cb,Kovchegov:2007pq,Spillane:2011yf,Sfetsos:1994xa,Albacete:2008vs,Khlebnikov:2010yt,Khlebnikov:2011ka,Chesler,Heller:2011ju,Albacete:2009ji,Aref'eva:2009kw,Aref'eva:2009wz,Wu,Albacete:2008ze, Taliotis:2009ne, Grumiller:2008va,Taliotis:2010pi}\footnote{These works refer to the $AdS_5$ geometry whose dual theory is the $\cal{N}$$=4$ sYM and not QCD. References \cite{CasalderreySolana:2011us,Edelstein:2009iv,Taliotis:2010kx} review the similarities and differences of the two gauge theories.  The works \cite{CasalderreySolana:2011us,Edelstein:2009iv,Bernamonti:2011vm,Janik:2010we} discuss the recent development in the field of applications of AdS/CFT in QCD. Particular emphasis in heavy-ion collisions  and the Quark-Gluon Plasma is given in \cite{Bernamonti:2011vm,Janik:2010we}.}.

 A direct outcome of this approach, is the estimate of lower bounds of the final multiplicity by using Penrose's idea of trapped surfaces.

 In this paper, we analyse  the formation of trapped surfaces in head-on collisions of shock waves in gravitational theories with more complicated bulk dynamics, and different types of vacuum solutions. An example are Einstein-dilaton theories with a scalar potential. Such theories have been argued, \cite{ihqcd,gubser}, to describe holographic  physics that is closer to QCD than the AdS theory. A phenomenological theory, Improved Holographic QCD (IHQCD)  has been constructed, \cite{gkmn,ihqcdrev}, that agrees well with both zero temperature and finite temperature YM data.

The idea is to explore how different aspects of the dynamics  affect the trapped surface that forms during the collision of  shock waves. The shock waves used are gravity (spin-2) shock waves. The different factors are the following:

\begin{itemize}

\item Different bulk geometries. There are several different possible geometries that that have been classified in \cite{ihqcd}.
    They are characterized  by their IR and UV (near-boundary) behavior.

    In the UV the typical behavior is asymptotically AdS. There is however interest in different non-AdS asymptotics, as they can capture the physics of the collision when the trapped surface forms mostly in the IR part of the geometry.

    The IR behavior  can be split into three large classes: (a) Confining geometries. (b) Non-confining geometries. (c) Unacceptable geometries (that violate the Gubser bound\footnote{This is a criterion introduced by Gubser in \cite{Gubser:2000nd} in order to test the acceptability of solutions with an IR singularity. It states that a naked singularity is acceptable (``good") if it can be covered by an infinitesimal horizon.
    The implications of this constraint for Einstein dilaton gravity were analyzed in \cite{ihqcd}.}). $AdS_5$ belongs to class (b). There is another special geometry that lies at the boundary of (a) and (b), the linear dilaton geometry.

\item Different profiles of the transverse distributions of energy. So far two types have been analyzed, uniform distributions or power-like (aka GPY, \cite{Gubser:2008pc}) profiles. We will add one more class namely exponential profiles, that are well localized in the transverse plane. Such energy profiles are closer to what the targets are in heavy-ion collisions. Except the uniform distributions, the other profiles are characterized by a length scale that controls the size of the energy distribution. Typically GPY profiles are fuzzier as they stretch to larger distances, while exponential profiles are sharply localized.

\item The option of cutting off the UV part of the bulk geometry, \cite{Gubser:2008pc}. This is motivated by the fact that we are striving to emulate the QCD behavior that is perturbative in the UV. In a perturbative regime, by definition,  multiplicities are small. It is in this regime that the geometry is expected to be unreliable in a holographic description. The simplest way to implement this is to stop the geometry at a point deemed to be the transition to the perturbative regime, and ignore the contributions above that point. This is not a controlled approximation, but it is expected to give useful hints.\footnote{In the bottom-up models of \cite{ihqcd} the gauge coupling constant, epitomized by the exponential of the dilaton is becoming small in the UV region. Despite this, the interactions of spin-2 matter are large in the UV. A potential way out is to advocate an asymptotically AdS metric in the string frame, but this cannot be done with an action with two derivatives, \cite{disec}.}

\end{itemize}

Our approach is to  solve  first the equations for the shock waves, in the presence of different types of metrics, with sources in the bulk.
Such solutions determine also the transverse distributions.
We then find the associated trapped surfaces, and the estimated lower bound on the generated entropy\footnote{An interesting approach for computing the entanglement entropy \cite{Casini:2011kv} and probing the scale-dependence of thermalization is provided by reference \cite{Balasubramanian:2011ur}.}. We will use the word entropy liberally in this paper to mean the area of the trapped surface. Strictly speaking this is a lower bound on the entropy, but it will always be obvious which entropy we are referring to. Moreover, entropy can be converted to multiplicity, and it is in this sense that we will use this two terms interchangeably.

Two characteristic examples are analyzed quantitatively and eventually compared to data.

\begin{enumerate}

\item The first is AdS$_5$, with a transverse distribution having a localized exponential profile and a cutoff in the UV at $r=1/Q_s$ with $Q_s$ the saturation scale (section \ref{UVQs}).
    We will name this setup ``AdS-Q$_s$" for future reference.

 \item     The second is a simplified metric emulating the IHQCD solution of \cite{ihqcd}, implementing confinement and asymptotic linear glueball trajectories, while it is asymptotically AdS$_5$ in the UV.
      In this case we have again a localized exponential profile in the transverse plane and no UV cutoff,  (section \ref{smM}).
      We will name this setup, with a slight abuse of language, ``IHQCD".

      \end{enumerate}

What we find is as follows:

\begin{itemize}

\item Both, our analytical calculations and our numerical analysis show that most of the entropy comes from the UV part of the geometry provided that the geometry reduces to the AdS$_5$ space near the boundary. This is not in contradiction with the naive expectation that in a strongly coupled theory the multiplicity comes from the high-energy part of the phase space available during the collision.

\item For uniform transverse distributions, the AdS$_5$ geometry, produces the least entropy among the geometries of section \ref{NoTD} .
    For non-confining geometries the entropy scales as $s^a$ for large s, with ${1\over 3}\leq a<{1\over 2}$ with $a={1\over 3}$ for AdS$_5$.
    For scaling confining geometries ${1\over 2} \leq a<1$. The case $a={1\over 2}$ corrected by logs corresponds to the geometries related to $IHQCD$.
    We have assumed above that the other geometries persist up to the boundary, or equivalently the trapped surface forms in the regime in which they are valid. These results are summarized in table \ref{T1}.
    However, as most realistic geometries are asymptotically AdS, the thrust of this result may be important only at intermediate energies, where IR geometries may take over.

    \item  For geometries that have a mass gap, discrete spectrum and confinement, the allowed transverse distributions has a spectrum of scales that is in one to one correspondence with the discrete spectrum of $2^{++}$ glueballs.
        The associated  entropy production, independently of transverse distribution, is less than AdS$_5$. It is assumed that all such confining geometries  are asymptotically AdS$_5$.

   \item An important puzzle of our analysis in the previous item is that the scattering of a distribution with transverse size associated with the lowest lying $2^{++}$ glueball does not seem to lead to a trapped surface.

\item     There are geometries with a UV energy-independent cut-off that lead to an asymptotic $\sim \log^2(s)$ behavior for the entropy (see table \ref{T2}).

        \item A general trend is that at equal total energy, the collision of distributions that have a larger transverse size leads to a larger entropy production.
          This implies that more dilute energy distributions  produce more entropy at fixed total energy.
          In particular a uniform transverse energy distribution produces (at equal total energy) more entropy that one with power-law or exponential transverse distribution. Similarly a power-law transverse distribution at the sane total energy  produces more entropy compared to an exponential transverse distribution with the same length scale.

   \item Comparing an AdS geometry, with an asymptotically AdS geometry that is confining in the IR we find that for all different types of transverse energy distributions that we have examined (uniform, power-law, or exponential), the AdS geometry generates substantially more entropy than the confining geometry, at the same (and large ) total energy, and transverse scale.

        \begin{table}[h!]\label{SGeom}
\begin{center}
\begin{tabular}{|l||*{15}{c|}} \hline
&$b(r)$ &${L\over r}$& ${L\over r} \exp[-\frac{r^2}{R^2}]$\\ \hline\hline
Transverse profile&&& \\ \hline\hline
Uniform &&$S_{unif}^{AdS}$&  $S_{unif}^{IHQCD}$  \\\hline
GYP &&$S_{GPY}^{AdS}$& Not studied \\\hline
Exponential &&$S_{exp}^{AdS}$&  $S_{exp}^{IHQCD}$\\\hline
 \end{tabular}
  \end{center}
  \caption{The several cases analysed and compared. Two geometries have been compared: A non confining geometry (the $AdS_5$) and a confining ($IHQCD$) one. None has a cutoff. The transverse profiles correspond to the cases of being constant, GPY (falling-off as a power law, \cite{Gubser:2008pc})  and falling-off exponentially.}
\end{table}

More to the point, table \ref{SGeom} defines the different contexts studied in this paper.
We find the following inequalities between the various trapped surface areas (keeping total energy and transverse size fixed)
\be
S^{AdS}_{unif}>S^{AdS}_{GPY}>S^{AdS}_{exp}\sp S^{IHQCD}_{unif}>S^{IHQCD}_{exp}
\ee
\be
S^{AdS}_{unif}>S^{IHQCD}_{unif}\sp S^{AdS}_{GPY}> S^{IHQCD}_{exp}\sp S^{AdS}_{exp}>S^{IHQCD}_{exp}
\ee

  \item In confining backgrounds, the entropy production increases as the confinement scale $\Lambda_{QCD}$ decreases, provided that the total energy and the transverse size are kept fixed.

  \item In the AdS-Q$_s$ setup the multiplicities grow with the atomic number almost linearly; in particular as $\sim A^{17/18}$ \cite{Collaboration:2011rta}.

  \item The lower bound on multiplicities for AdS-Q$_s$ is given in equation (\ref{cfS}). For the IHQCD setup it is given in for the (\ref{fitc}).
      Both formulae once fit to RHIC data, make the same (correct) prediction $2.76$ TeV PbPb LHC data, \cite{Collaboration:2011rta}.
      A single parameter is used in these fits, namely the overall constant coefficient of the leading (large) s-dependence.

\end{itemize}

The implications of the conclusions above do not a priori apply verbatim to QCD because:

\begin{enumerate}

\item We are discussing glueballs only, and not nuclei.

\item The holographic backgrounds used are approximations to various gauge dynamics. Only AdS is an exact background for sYM, but all backgrounds,  including AdS used for QCD,  are approximations.

\item The final entropy produced may be and is usually larger than the trapped surface area.
\end{enumerate}

The first issue may not be  of prime importance at high energy. The reason is that the energy released at mid-rapidity in heavy-ion collisions is expected to be mostly gluons. Quarks will provide corrections to this but they are not expected to change this picture drastically.

The second issue is of importance. Although by now bottom-up models of YM can provide reliable calculations matching lattice calculations, their structure in the UV is more shaky. It is not a priori clear where the geometric description breaks down, and such transitions may be at different places for different observables.

The third issue is also important for quantitative predictions.
There are very few cases where the final entropy has been calculated numerically in the collision of shock waves, \cite{Chesler,Wu}.
In \cite{Wu} in particular it was shown that the released entropy at the end of the collision process is 60\% larger that the bound found from the trapped surface calculation. Moreover conformal invariance in AdS makes this percentage to be independent of the collision energy.

This suggests that for collisions and bulk geometries where the majority of the     trapped surface area come for the UV, AdS part of the geometry, we should still expect that the relative factor relating the final entropy release to the area of the trapped surface to be almost energy-independent.

The analysis in this paper is providing important and potentially general clues on multiplicity generation in high energy collisions using holography.
It must be however be backed-up by a more reliable calculation of the gravitational evolution, but we leave this for future work. Another interesting issue is whether there are differences between the high-energy scattering of $2^{++}$  glueballs studied in this paper and $0^{++}$ glueballs. Although we do not have reasons to expect major differences, an analysis in this case should be done.

We organize this write-up as follows: We begin in section \ref{ED} by showing the way one may build more realistic geometries that come closer to QCD. This is achieved by considering scalar gravity. In section \ref{GTS}, we state the equations that compute the entropy production of the collided shock-waves and hence of the gauge matter indicating that the calculation reduces to a boundary valued problem. In section \ref{SG} we present the subclass of geometries we investigate. Sections \ref{NoTD} and \ref{NUTD} refer to various geometries with uniform or non-uniform transverse dependence respectively. The entropy of the trapped surface is computed for each case. Section \ref{RUV} deals with the simplest way of removing the weak coupling entropy production: At higher energy scales,  where coupling is weak, the contribution to the entropy production should be less important. In the geometrical language of AdS/CFT this would imply that one should modify the way the fifth coordinate of AdS/CFT at the UV is treated.  In section \ref{RS}, we present our results and make our predictions. Finally, in section \ref{finally}, we comment on various aspects of our investigations and conclusions. We omit intermediate details of our calculations for the three appendices at the end. In particular, appendix \ref{A} serves as a practical introduction to the theory of trapped surfaces. Appendix \ref{B} shows how one may localize the (five-dimensional) bulk sources and appendix \ref{C} proves a useful equation that is needed for subsection \ref{Ltd}.

\section{Einstein-Dilaton gravity}\label{ED}
%%%%%%%%%%%%%%%%%%%%%%%%%%%%%%%%%%%%%%%%%%%%%%%%%%%%%%%%%%%%%%%%%%%%%%%%%%%%%%%%%%%%
We start from the action:
\begin{equation}
   S_5=-M^3\int d^5x\sqrt{g}
\left[R-{4\over 3}(\partial\Phi_s)^2+V(\Phi_s) \right]
    \label{app1}\end{equation}
 where $\Phi_s$ is the scalar field dual to the YM coupling constant.
We first find shock wave solutions in this theory of the form
\be\label{1shgmn}
ds^2=b(r)^2\left[{dr^2}+dx^idx^i-2dx^+dx^-+\phi(r,x^1,x^2)\delta(x^+)(dx^+)^2\right], \qquad \Phi_s = \Phi_s(r,x^+)
 \ee
with the asymptotically AdS boundary  at $r=0$. Compatibility of these equations implies that $\partial_{+}\Phi=0$. Eliminating $\Phi_s$ using the equations of motion, the equation for $\phi$ is

\begin{align}\label{deF}
\left(\nabla_{\perp}^2+3\frac{b'}{b}\partial_r+\partial_r^2\right)\phi =-2\kappa_5^2J_{++}, \hspace{0.15in}\nabla_{\perp}^2\equiv \partial_i\partial_i,
 \hspace{0.15in}b'\equiv \partial_rb(r)    \hspace{0.15in} \kappa_5^2 \equiv 8 \pi G_5
\end{align}
where we have introduced a stress-tensor $J_{++}$.

\section{Trapped surfaces}\label{GTS}
%%%%%%%%%%%%%%%%%%%%%%%%%%%%%%%%%%%%%%%%%%%%%%%%%%%%%%%%%%%%%%%%%%%%%%%%%%%%%%%%%%%%%%%
Trapped surfaces are created when two shocks like the one of (\ref{1shgmn})\footnote{We call it $\phi_1$ to distinguish it from the second that moves along $x^+$ that we call  $\phi_2$. In our case the shocks will be taken identical and hence the subscripts will be soon dropped.} which moves along $x^-$  collide\footnote{A more complete set of notes on the theory of trapped surfaces may be found in appendix \ref{A}.}. In terms of metrics before the collision, one then has

\begin{align}\label{gmn2f}
ds^2=b(r)^2\Big[   {dr^2}+dx^idx^i&-2dx^+dx^- + \phi_1(r,x^1,x^2)\delta(x^+)(dx^+)^2\\ \notag &
+\phi_2(r,x^1,x^2)\delta(x^-)(dx^-)^2  \Big], \hspace{0.2in}x_{\pm}<0.
 \end{align}

Associated with the shock-wave $\phi_1$ in (\ref{1shgmn}),  we parametrize (half of the) trapped surface $S_1$ by
\begin{align}\label{TS}
x^+=0  \hspace{0,2in} x^-+\frac{1}{2}\psi_1(x^1,x^2,r)=0
\end{align}
where $\psi_1$ remains to be determined. It is also useful to rescale the functions $\phi_1$ 
and $\psi_1$ by defining
\begin{align}\label{Psi}
\Phi_1=b(r)  \phi_1     \hspace{0,2in} \Psi_1= b(r) \psi_1.
\end{align}
$\Psi_1$ satisfies the following differential equation
\begin{align}\label{1box1}
 (\Box_{AdS_3}-A) (\Psi_1-\Phi_1)=0 \hspace{0.2in}A \equiv \frac{\partial_r(b(r) b'(r))}{b(r)^4}
\end{align}
where $\Box_{AdS_3}$ is defined with respect to the metric
\begin{align}\label{1AdS3}
ds^2= b(r)^2  \left( dx^2_{\perp}+dr^2\right), \hspace{0.2in} dx^2_{\perp} \equiv (dx^1)^2+(dx^2)^2.
\end{align}
We point out that once (\ref{deF}) is solved,  then $(\Box_{AdS_3}-A(b(r))) \Phi_1$ provides a source term for  $(\Box_{AdS_3}-A(b(r))) \Psi_1$. The missing ingredient is the boundary conditions that are given by
 \begin{align}\label{1BC}
\Psi_1\Big |_C=0 \hspace{0.4in}  \frac{1}{b(r)^2} \sum_{i=1,2,r}\left[ \nabla_i \Psi_1 \nabla_i \Psi_1 \right]  \Big |_{C}=8
 \end{align}
 for some curve $C$ which defines the boundary of the trapped surface and where both, $S_1$ and (the associated surface to $\phi_2$,) $S_2$ end. The entropy is then bounded below by the area of the surface obtained by adjoining the two pieces of the trapped surface associated with each of the shocks as
\begin{align}\label{ent1}
S \geq S_{trap}=2\times \frac{1}{4G_5} \int_C \sqrt{det|g_{AdS_3}|}dr d^2x_{\perp}=\frac{\pi}{2G_5} \int_{r_{C_2}}^{r_{C_1}}  b(r)^3 x^2_{\perp}(r)
\end{align}
where the (generalized) curve $C$ defines the boundary of the trapped surface $S_1$ and $S_2$ which are identical; thus the overall factor of 2. The integral with respect to the transverse coordinates gives $x^2_{\perp}(r)$ when considering a head-on collision. Typically, $r_C{_1}$, $r_{C_2}$ and $x_{\perp}$ carry the information of the shock $\phi$ \footnote{We drop the subscripts $1,2$ from $\phi$'s and $\psi$'s from now on.}.
We conclude by pointing out that the un-scaled version of the trapped surface (see (\ref{Psi})) for a head-on collision is given by
\begin{align} \label{psik1}
     \left(\nabla_{\perp}^2+3\frac{b'}{b}\partial_r+\partial_r^2\right) (\phi-\psi)=0, \hspace{0.2in}   \psi=0\big|_C \hspace{0.01in}, \hspace{0.2in} (\partial_r \psi)^2+(\partial_{x_{\perp}} \psi)^2\big|_C=8
 \end{align}
 while the entropy is still given by (\ref{ent1}). In the absence of transverse dependence one ignores $x^2_{\perp}$ from (\ref{ent1}) which measures entropy/transverse area in this case.
%%%%%%%%%%%%%%%%%%%%%%%%%%%%%%%%%%%%%%%%%%%%%%%%%%%%%%%%%%%%%%%%%%%%%%%%%%%%%%%%%%%%%%
\section{Shock geometries}\label{SG}

We have analyzed a number of bulk geometries by examining a class of different scale factors, that classify the non-conformal behavior of Eistein-dilaton gravity models. In terms of their behavior in the far IR ($r\to \infty$, or $r\to r_0$), we have the following cases \cite{ihqcd,gkmn}:
\begin{enumerate}

\item $b\sim r^{a}$,
with $a \leq -1$. The AdS case corresponds to $a=1$. This corresponds to quasiconformal geometries, with no confinement, continuous spectrum and a mass gap, with potential asymptotics as $\Phi_s\to\infty$, $V\sim e^{Q\Phi_s}$, $Q<{4\over 3}$.

\item Confining backgrounds that are scale invariant in the IR, \cite{GK}, with $b(r) \sim (r_0-r)^a$, $a>{1\over 3}$. In this case $r_0$ is finite and signals the position of an IR singularity that satisfies the Gubser bound for $a>{1\over 3}$. They have a discrete spectrum of glueballs and a mass gap. The potential asymptotics as $\Phi_s\to\infty$ are $V\sim e^{Q\Phi_s}$, $Q>{4\over 3}$.

 \item Confining backgrounds with $b(r)\sim e^{-(\Lambda r)^a}$, $a>0$.
 They have a discrete spectrum and a mass gap. The potential asymptotics as $\Phi_s\to\infty$ are $V\sim e^{{4\over 3}\Phi_s}~\Phi_s^{a-1\over a}$.

      \item Confining backgrounds with $b(r)\sim e^{-\left({\Lambda \over r-r_0}\right)^a}$, $a>0$ and $r_0$ as in the second point above. They have a discrete spectrum and a mass gap. The potential asymptotics as $\Phi_s\to\infty$ are $V\sim e^{{4\over 3}\Phi_s}~\Phi_s^{a+1\over a}$.

\end{enumerate}

\section{Shocks with uniform transverse space dependence}\label{NoTD}
%%%%%%%%%%%%%%%%%%%%%%%%%%%%%%%%%%%%%%%%%%%%%%%%%%%%%%%%%%%%%%%%%%%%%%%%%%%

In this case the shock $\phi(x^+,r)$ can be determined and the trapped surface ends at $r=r_H$ with $r_H$ determined from\footnote{The delta function in (\protect \ref{10}) may be in principle replaced by any function of $x^+$.}
\be
\Phi(r,x^+)=E\delta(x^+)\int{dr\over b^3}
\sp {b^3(r_H)}={E\over \sqrt{8}}
 \label{10}
\ee
with $E\sim s^{1\over 2}$, and $s$ the center of mass energy of the collision.
The area of the trapped surface is
\be
A_{trap}\simeq \int_{\infty}^{r_H}b^3 dr
\ee
We may therefore estimate the energy dependence of the trapped area for different bulk geometries.
\begin{itemize}

\item For non-confining scaling theories $b\sim r^{-a}$,
with $a \geq 1$
we obtain
$A_{trap}\sim s^{3a-1\over 6a}$. The AdS case corresponds to $a=1$. This agrees with previous estimates for the AdS case, \cite{Gubser:2008pc}.

\item Confining backgrounds with $b(r) \sim (r_0-r)^a$, $a>{1\over 3}$. In this case we obtain $A_{trap}\sim s^{3a+1\over 6a}$ at high energy. The exponent varies between ${1\over 2}$ and 1.

 \item Confining backgrounds with $b(r)\sim e^{-(\Lambda r)^a}$. In this case $A_{trap}\sim s^{1\over 2}(\log s)^{a+1\over a}$ at high energy.

      \item Confining backgrounds with $b(r)\sim e^{-\left({\Lambda \over r-r_0}\right)^a}$. In this case $A_{trap}\sim s^{1\over 2}(\log s)^{1-a\over a}$ at high energy.

\end{itemize}

Taking into account that $S_{trap}\sim A_{trap}$ we conclude that in all cases the entropy production is larger than AdS at high enough energies.
%%%%%%%%%%%%%%%%%%%%%%%%%%%%%%%%%%%%%%%%%%%%%%%%%%%%%%%%%%%%%%%%%%%%%%%%%%%%%%%%%%%%%%%
\section{Non-uniform transverse dependence}\label{NUTD}
%%%%%%%%%%%%%%%%%%%%
We assume the shock(s) have non-trivial transverse dependence and we solve the homogeneous equation (\ref{deF}) by separating variables. We obtain the following set of differential equations (assuming rotational symmetry on the transverse plane)
\begin{align}\label{SV}
\phi_k\sim f_k(x_{\perp})g_k(r)  \hspace{0.2in}   \left( \partial_{x_{\perp}}^2+\frac{1}{x_{\perp}}\partial_{x_{\perp}}-k^2   \right)f_k(x_{\perp})=0
\hspace{0.2in}   \left( \partial_{r}^2+3\frac{b'(r)}{b(r)}\partial_{r}+k^2   \right)g_k(r)=0.
\end{align}
The first equation yields
\begin{align}\label{Bes}
f_k(x_{\perp})=C_1 K_0(k x_{\perp})+C_2 I_0(k x_{\perp})
\end{align}
while the solution to $g(r)$ depends on the scale factor $b(r)$. Strictly speaking, the solutions corresponding to $K_0$ do not exactly solve the homogeneous differential equation for $f_k$. They induce a delta function source and they satisfy
\begin{align}\label{nhom}
\left(\nabla_{\perp}^2+3\frac{b'}{b}\partial_r+\partial_r^2\right)K_0(k x_{\perp})g_k(r)=-2\pi \delta^{(2)}({\bf x_{\perp}})g_k(r).
\end{align}
It is evident that $J_{++}$ from (\ref{deF}) consists from an appropriate linear combination of $K_0(k x_{\perp})g_k(r)$'s. In the remaining of this section we analyze several $b(r)$'s.

\subsection{Power-like transverse distributions}

We will study solutions to (\ref{SV}) where the transverse distributions fall-off as a power in the transverse plane. Such solutions were considered for AdS in \cite{Gubser:2008pc}. In the remaining of this section we analyze several different types of scale factors $b(r$) and we derive the boundary that specifies the trapped surface as a function of the energy (see equations (\ref{Eq}) and (\ref{cond1})). Then using (3.6) one may compute the entropy bound S; the final answers for S are summarized in section 8 (see tables \ref{T1},  \ref{T2},  \ref{T3}).

\subsubsection{Power-law $b(r)$}\label{bpl}
%%%%%%%%%%%%%%%%%%%%%%%%
We begin by studying the power-like scale factor
\begin{align}\label{pl}
b(r)=\frac{1}{L^a}(r-r_0)^a.
\end{align}
When $a>{1\over 3}$ and $r_0$ finite, this corresponds to  a confining geometry, with IR singularity at $r=r_0$, \cite{ihqcd}. This can be resolved by uplifting to higher dimensions, \cite{GK}.
It can also include non-confining geometries, if $r_0=0$ and $a<-1$.
$a=-1$ corresponds to AdS. The boundary for $a>1/3$ is at $r\to\infty$ while for $a<0$ is at $r \to 0$.

Assuming a point-like $J_{++} =E \delta(r-r')\delta(x^1)(x^2)\delta(x^+)$ ($r'$ is the $r$ coordinate of the source) in (\ref{deF}), the shock $\phi$ is given by
\begin{align}\label{ph}
\phi=(\frac{r-r_0}{L})^{-\frac{1}{2}(1+3 a)} \tilde{\Phi}(q),  \hspace{0.2in }   \tilde{\Phi}(q)=\frac{P^{\frac{1}{2}}_{-1+\frac{3}{2}a}(1+2q)}{(q(1+q))^{\frac{1}{4}}},
 \hspace{0.2in}q=\frac{x_{\perp}^2+(r-r')^2}{4(r-r_0)(r'-r_0)}
\end{align}
where $P^{\mu}_{\nu}$ is the associated Legendre polynomial and where the tilde on $\tilde{\Phi}$ is  to remind us that the rescaling of $\phi$\footnote{And likewise, the rescaling of $\psi$.} is not by a factor of $b(r)$ to the first power (see (\ref{Psi})) and will be omitted from now on.
The presence of the combination $q$, natural in AdS space, \cite{Gubser:2008pc} for more general power-like metrics, can be explained,  as such metrics are conformally AdS, and can be lifted to AdS metrics in a higher dimensional
 space-time, \cite{GK}.

 The equation for the trapped surface satisfying the first boundary condition is given by
\begin{align}\label{PSIolan}
\Psi(q)=\Phi(q)-\frac{\Phi(q_C)}{\Phi_-(q_C)}\Phi_-(q), \hspace{0.2in }  \Phi_-(q)=\frac{Q^{\frac{1}{2}}_{-1+\frac{3}{2}a}(1+2q)}{(q(1+q))^{\frac{1}{4}}}
\end{align}
where $\Phi_-$ is the solution to the homogeneous differential equation satisfied by $\Phi$ which is finite at $q=0$. $Q_{\nu}^{\mu}$ is the associated Legendre polynomial of the second kind while $q_C$ denotes the value of $q$ at the boundary of the trapped surface. The second boundary condition in (\ref{psik1}) yields
 \be\label{Eq}
   \left(\frac{L}{r_C-r_0}\right)^{\frac{3}{2}(1+ a)} \frac  {E\kappa_5^2 (r'-r_0)}{L^3} \sim q_C(1+q_C)\Phi_-(a;q_C)\rightarrow
 \left\{ \begin{array}{ll}
q_C^{\frac{1}{2}(1+3a)}, & a>{1\over 3},\\ \\
 q_C^{-\frac{3}{2}(a-1)} ,& a\leq-1.
\end{array}\right.
\ee
where the quantity on the right of the arrow denotes the high energy limit. We note the agreement with \cite{Gubser:2008pc} when $a=-1$, $r_0=0$ and $r'=L$. The $S_{trap}$ is given by
\begin{subequations} \label{fullS}
 \be
S_{a>\frac{1}{3}} \sim \pi \int_ {r_{C_1}(E)}^{r_{C_2}(E)} (r-r_0)^{3a} \left(4 (r-r_0)(r'-r_0)\left(\frac{E}{(r-r_0)^{\frac{3}{2}(a+1)}}\right)^{\frac{2}{1+3a}}
-(r-r')^2) \right) dr ,\label{fullSa+}
\ee
\be
S_{a \leq-1} \sim \pi \int_ {r_0}^{r_{C_2}(E)} (r-r_0)^{3a} \left(4 (r-r_0)(r'-r_0)\left(\frac{E}{(r-r_0)^{\frac{3}{2}(a+1)}}\right)^{\frac{-2}{3(a-1)}}
-(r-r')^2) \right) dr.   \label{fullSa-}
\ee
\end{subequations}
The limits of integration are found from (\ref{Eq}) when $x_{\perp}=0$. Performing the integrations and keeping the leading contribution in $E$ taking into account the range of $a$ for each case, we finally arrive at
\begin{subequations} \label{fullSEAll}
 \be
S_{a>1/3} \sim E^{\frac{3a+3}{3a+2}}\label{fullSEap}
\ee
\be
S_{a\leq-1} \sim E^{\frac{(3a+1)}{3 a}}  .  \label{fullSEam}
\ee
\end{subequations}
As a cross check we note that (\ref{fullSEam}) for $a=-1$ reproduces the result of \cite{Gubser:2008pc}.
 We remark that the trapped surface for $a>1/3$ exists because $r_0$ is finite.

%%%%%%%%%%%%%%%%%%
\subsubsection{Exponential $b(r)$}
We will take the scale factor b(r) in this case to be
 \begin{align} \label{bex}
b=e^{-r/R}.
\end{align}
This corresponds to a marginal confining geometry, that of a linear dilaton, \cite{ihqcd}.
%%%%%%%%%%%%%%%%%%%%%%%%%
Rescaling $ \phi=e^{\frac{3 r}{2 R}} \Phi(u(r,x_{\perp}))$\footnote{And likewise, rescaling $\psi$.} and assuming a point-like $J_{++}$ as in the power-low case, we obtain
\begin{align}\label{exsol}
 \Phi(u)\sim \frac{E k_5^2}{L}\frac{1}{\sqrt{u}} e^{-\frac{3}{2 }\sqrt{u}}, \hspace{0.2in} u \equiv \frac{(r-r')^2+(x_{\perp}-x_{\perp}')^2}{R^2}.
 \end{align}
where $r'$ is the $r$ coordinate of the source. The trapped surface equation is
\begin{align}
 \Psi(u)=\Phi(u)-\frac{\Phi(u_C)}{\Phi_-(u_C)}\Phi_-(u) \hspace{0.2in} \Phi_-(u)=\frac{\sinh(\frac{3}{2 }\sqrt{u})}{\sqrt{u}}
  \end{align}
while the rescaled boundary condition $\frac{4}{R^2} e^{3r/R} u(\Psi'(u))^2\big|_C=8$ finally yields
  \begin{align}\label{cond1}
E \sim \frac{L^3}{\kappa_5^2 R} e^{-3r_C/2R} \sqrt{u_C}  \sinh(\frac{3}{2 }\sqrt{u_C})
  \end{align}
where the quantity on the right of the arrow denotes the high-energy limit. The trapped surface is found by working as in subsection \ref{bpl}. In this case also, the trapped surface exists either when $r$ is allowed to extend in the interval $[-\infty,\infty]$ or when a UV cut-off is placed (see discussion below (\ref{fullS})). 

Allowing for $r$ to take negative values, so that the geometry is extended towards the boundary, the entropy is computed from (\ref{cond1}) numerically: One, solves this equation with respect to $x_{\perp}^2=x_{\perp}^2(r,E)$ and integrates  as in (\ref{ent1}). The two limits of integration are found numerically from (\ref{cond1}) again by setting $x_{\perp}=0$. The analysis shows that
\begin{align}\label{SnumEx}
 S_{trap} \sim s^{1.66}\log^{1.17}(s).
  \end{align}
The result indicates a larger growth of $S_{trap}$ than the uniform profile case (see third bullet of section \ref{NoTD}). This, does not contradict our general conclusions as here the exponential metric is assumed to be valid up to the UV boundary.
There is a way to derive a simple formula for $S_{trap}$ when a UV cut-off (see section \ref{RUV}) is placed. If for instance, we cut the surface at some $r>0$ then equation (\ref{cond1}) implies that as $E$ increases,  $u_C$ increases logarithmically: $\log^2(E) \sim r_C^2+x_{\perp C}^2$. Hence, equation  (\ref{ent1}) yields
\begin{align}\label{ScutEx}
 S_{trap} \sim x_{\perp C}^2\sim \log^{2}(s).
  \end{align}
  This result looks similar with Froissart bound estimates for cross sections \cite{Kang:2004jd,Giddings:2002cd}. We do not know if there is a connection.
 %
%%%%%%%%%%%%%%%%%%%%%%%%%%%
\subsection{Localized transverse distributions}\label{Ltd}
%%%%%%%%%%%%%%%%%%%%%%%%%%%%%%

We built a single-mode shockwave $\phi_k$ out of the solutions of (\ref{SV}) demanding to be square integrable for all $x_{\perp}$. The result is
\begin{align} \label{phik}
\phi_k=\frac{E\kappa_5^2}{4\pi L^3k^2}g_1(kr)K_0(k x_{\perp})\delta(x^+), \hspace{0.1in} \mbox{          assuming       } \hspace{0.1in}  g_1(k r)\approx k^4r^4\Big|_{kr \ll1}
\end{align}
where E is the energy carried by the shock. $g_{1,2}(r)$, the radial solutions,  are dimensionless, because $\phi_k$ has dimensions of length. The overall constant ensures that the gauge stress tensor,  which according to the AdS dictionary is given by
\begin{align} \label{Adsdic}
T_{++}=\frac{2L^3}{\kappa^2_5}\lim_{r \to 0}\frac{\phi_k}{r^4},
\end{align}
when is integrated in space, it yields the total energy $E$. The trapped surface is defined by
\begin{align} \label{1psik2}
     \left(\nabla_{\perp}^2+3\frac{b'}{b}\partial_r+\partial_r^2\right) (\phi_k-\psi_k) \hspace{0.2in}   \psi_k=0\big|_C \hspace{0.2in} (\partial_r \psi_k)^2+(\partial_{x_{\perp}} \psi_k)^2\big|_C=8
 \end{align}
and yields
\begin{align} \label{psik}
\psi_k= \phi_k(r,x_{\perp })-\frac{  \sum_{k'} \left(C_{k'}^1g_1(k' r )+C_{k'}^2g_2(k' r) \right)I_0(k' x_{\perp})  } {   \sum_{k'} \left(C_{k'}^1g_1(k' r_C )+C_{k'}^2g_2(k' r_C) \right)I_0(k' x_{\perp C})  }  \phi_k(r_C,x_{\perp C }).
\end{align}
for some coefficients $C_k$ to be determined. The $K_0$'s do not participate as they induce source terms\footnote{In appendix \ref{B} it is shown how these sources may be localized.}.
If we assume that at small $r$, $g_1 \rightarrow r^4 ~\Pi(r)$ with $\Pi(r)$ a regular function  with at least one real root, and that $g_1$ has no multiple roots then, as it is
proved in appendix \ref{C}, only the coefficient $C^1_k$ is non-trivial. This assumption turns out to be true for the cases we are studying in this paper. The boundary condition for the trapped surface then satisfies
\begin{align}  \label{BCk3}
\left( \frac{E\kappa_5^2}{4\pi L^3 k}\right)^2 \left(  \frac{g_1(k r)}{I_0(k x_{\perp}) k x_{\perp}}  \right)^2 \Big|_C=8
\end{align}

The high-energy limit of the trapped surface is defined by
\begin{align}  \label{HELI}
\frac{E\kappa_5^2}{4\pi L^3 k} \gg1.
\end{align}
(see equation (\ref{HELI})). Equation (\ref{BCk3}) defines the boundary of the trapped surface as a function of the energy $E$ and the wavenumber $k$. We will find specific examples where there exist $r_{C_1}=0$ and  $r_{C_2}$ for suitably chosen $b(r)$ and $k$ in subsection \ref{ds}.

%%%%%%%%%%%%%%%%%%%%%%%%%%%%%%
 \subsubsection{Discrete spectra}\label{ds}
 %%%%%%%%%%%%%%%%%%%%%%%%%%%%%
We consider a confining  scale factor $b$ which asymptotes to $AdS$ in the UV and behaves as in IHQCD in the IR,
\begin{align} \label{e^r2}
b(r)=\frac{L}{r}e^{-\frac{r^2}{R^2}}
\end{align}
Although the precise solution, that fits YM data is slightly different, \cite{data}, we will use the one in (\ref{e^r2}) as it does not affect the high energy asymptotics. This solution has been discussed in appendix G of \cite{ihqcd}. The length scale $R$ is the analogue of the QCD scale.  We may parametrize it as $R\sim \Lambda_{QCD}^{-1}$.

Unlike non-confining cases, in confining bulk geometries the physical spectrum of glueballs, namely the fluctuations of the metric and the scalar dilaton, is discrete, \cite{ihqcd}.
The radial equations that define both the shockwave profile, in equation (\ref{deF}), as well as the equation for the trapped surface, (\ref{psik1})
involve the same radial equation as the one that determines the spectrum of spin-2 fluctuations ($2^{++}$ glueballs).

When it comes to determine the presence of a trapped surface, a similar condition appears: there is no trapped surface unless the solutions to (\ref{psik1}) are normalizable (see (\ref{norm})). This is directly obvious in shockwaves determined by a single graviton wavefunction and can be shown in general.

As normalizability of the wavefunction is important for the existence of the trapped surface, the spectrum of shockwave profiles is discrete, with the transverse momentum determined from the $2^{++}$ glueball masses $m_n$ by $|k_n|=m_n$.
Therefore, the transverse profiles of the shockwave distributions, not surprisingly, are determined by the normalizable $2^{++}$ (graviton) wavefunction.

For the metric (\ref{e^r2}), when $k$ takes the particular subset of values
 \begin{align} \label{kn1}
k_n^2= (n+2) \frac{12}{R^2} \hspace{0.2in} n = 0,1,2,3,...
\end{align}
 the radial solution of  equation (\ref{SV}) reduces to a (finite) polynomial that behaves as $r^4$ at small $r$ and is normalizable. Normalizability in the radial direction is defined when the corresponding eigenfunction $g_1(k_n r)$ satisfies
\begin{align} \label{norm}
\int b(r)^3 |g_1(k_n r)|^2 dr<\infty.
\end{align}
The set of the normalizable eigenfunctions is given by
\begin{align} \label{Lp}
\frac{r^4}{R^4}L^{(2)}_n(3 r^2/R^2), \hspace{0.2in} n=0,1,2,...
\end{align}
where $L^{(2)}_n$ are the (finite) associated Laguerre polynomials of degree $n$. This is the qualitative behavior in any background that is confining with a discrete spectrum of glueballs and a mass gap. The values in (\ref{kn1}) coincide with the mass-spectrum of $2^{++}$ glueballs.

%%%%%%%%%%%%%%%%%%%%%%%%%%%%%%%%%%%%%%%%%%%%%%
  \begin{figure}[h!]
\centering
\includegraphics*[scale=0.8,angle=0,clip=true]{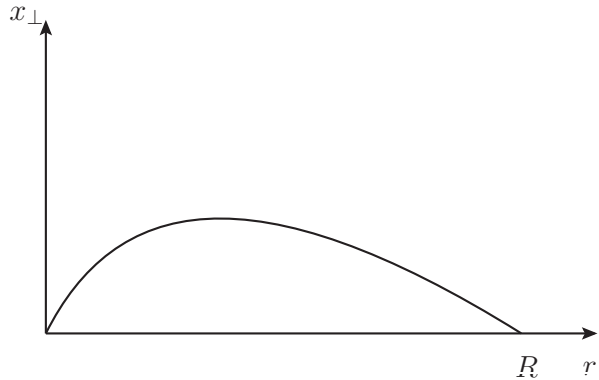}
\caption{The (closed) boundary $x_{\perp C}(r_C,E)$ of the trapped surface for the first excited state corresponding to $b(r)=L/r e^{-r^2/R^2}$ for fixed energy $E$. There is a family of such curves; one for each value of $E$ which moves upwards for larger $E$. The section $ x_{\perp}=0$ is a part of the boundary as should because the source has coordinate ${\bf x_{\perp}=0}$. The surface does not extend beyond $r>R$ which implies that there is no entropy production for energies greater than $1/R$, a scale which  is naturally identified with $\Lambda_{QCD}$.}
\label{TSB}
\end{figure}
%\vspace{-0.3in}
%%%%%%%%%%%%%%%%%%%%%%%%%%%%%%%%%%%%%%%%%%%%%%%%

 We now analyze the example  (\ref{e^r2})  and study the formation of a trapped surface via collision of such quantized transverse distributions.  The case $n=0$ gives a shock that behaves as $g_1\sim r^4$ for all $r$ and hence it has only a single root (see discussion after (\ref{HELI}). We therefore conclude that this ground state mode will produce no trapped surface  and therefore  no thermal medium.

We next  considering the $n=1$ case which yields the two independent solutions
\begin{align} \label{gi}
g_1(k_1 r)& = (36)^2 \frac{ r^4}{R^4} \left(1-\frac{r^2}{R^2}\right)  ,  \notag\\
g_2(k_1 r)& =\frac{ r^4}{R^4} \left(1-\frac{r^2}{R^2}\right)     \left(- e^{3\frac{r^2}{R^2}} \frac{ R^2 (-9 r^4 + 6 r^2 R^2 + R^4)}{r^4 \left(R^2-r^2\right) }+27E_i(3\frac{r^2}{R^2})    \right).
\end{align}
The solution $g_1$ has the right asymptotics in order to generate a closed trapped surface for all energies. The shock given by $g_1$ has $r^4$ behavior at small $r$ in accordance with the expectation value of the gauge theory stress-energy tensor. The boundary along the $r$ direction is given by $r_{C_1}=0$ and  $r_{C_2}=R$. Indeed, when $r$ takes these two values, the numerator of the left-hand-side of (\ref{BCk3}) becomes zero and compensates the zero of the denominator when $x_{\perp}=0$ as figure \ref{TSB} depicts.

 The (two pieces of the transverse) exponential profile trapped surface (without any cut-offs) yield
\begin{subequations} \label{Skfull}
 \be
S_{trap}^{k_n} =\frac{ L^3}{\kappa_5^2} \frac{8 \pi^2 }{12(n+2)}\int_0^{y_0} \frac{e^{-3 y^2}}{y^3}\frac{ \tilde{x}_{\perp}^2}{2}(n,E R;y) dy,
\ee
\be
   \frac{ER \kappa_5^2 \big|g_1(y\sqrt{{12(n+2)}})\big|}{8\pi L^3 \sqrt{24(n+2)}}=\tilde{x}_{\perp}I_0 (\tilde{x}_{\perp})
\ee
\end{subequations}
where $y_0$ is the highest root of $g_1(y\sqrt{{12(n+2)}})=0$ and  $n$ denotes the $n^{th}$-excitation. For the case at hand, $n=1$
while $g_1$ is given by (\ref{gi}) with $y=r/R$.

It is evident that the entropy, unlike \cite{Gubser:2008pc}, depends not only on the transverse size $R /\sqrt{12(n+2)}$ but in addition on the the confinement scale $1/R$. In fact, these two parameters may be varied independently.

We may compare with the analogous AdS calculation where the transverse profile is taken to be the same as here, with characteristic scale $k$. The confining theory has another scale $1/R$ that can be traded with varying the integer $n$. We find that for any $n$ the area of the trapped surface in AdS is always larger than that in the confining background.

  %%%%%%%%%%%%%%%%%%%%%%%%%%%%%%%%%%%%%%%%%%%%%%%
%%%%%%%%%%%%%%%%%%%%%%%%%%%%%%%%%%%%%%%%%%%%%%%%%%%%%%%%%%%%%%%%%%%%%%%%%%%%%%%%%%%%%%
\section{Accommodating asymptotic freedom}\label{RUV}
%%%%%%%%%%%%%%%%%%%%%%%
As was pointed out in \cite{Gubser:2009sx}, the UV (small $r$) should not contribute importantly to the entropy production $S$.
The reason is that by definition for perturbation theory to be valid, the generated particle multiplicities must be small. Many examples are known when large multiplicities imply the breakdown of perturbation theory, with the sphaleron case the most prominent one, \cite{bachas}.

We expect that in the QCD UV, at some point perturbation theory takes over, and this is defined as the regime in which the generation of entropy is small compared to that generated from lower scales. This transition we will approximate as an abrupt transition: we will assume that this is a radial position $r_{UV}$, below which we can use the gravitational description, while above it standard perturbation theory takes over. We will neglect the perturbative contribution to the entropy as we expect it to be small.

 Therefore in this approximation we will introduce a UV cutoff $r=r_{UV}$ in the trapped surface  that will simulate the emergence of weak coupling in the UV. The position of $r_{UV}$ must be determined, and at this stage it appears as an additional phenomenological parameter.

A related question is to what extend geometries with varying coupling constant
like IHQCD implement the fact that interactions are weak near the UV, as we would have expected from QCD. The answer is that in asymptotically AdS backgrounds, even as the string coupling  $e^{\phi}\to 0$ in the UV, the graviton interactions remains strong. It would have been probably different if the geometry becomes AdS in the string frame. However in this case, (a) the boundary geometry is singular (b) Such a case cannot be a solution to a gravitational  action with two derivatives only.

In the sequel we will cutoff the geometry in the UV and  explore the result.

\subsection{Energy-independent cut-off}

The work \cite{Gubser:2009sx} suggests that in the high-energy limit, the entropy,  for the geometries studied, should be given by
\begin{align} \label{Sex1}
S_{trap}(E) \sim \int_{r_{C_1}(E)}^{r_{C_2}(E)}b(r)^3x^2_{\perp}(E \rightarrow \infty, r)dr, \hspace{0.2in}r_{C_1}(E)\rightarrow C_1
\end{align}
where $C_1$ is an energy independent constant\footnote{In \cite{Gubser:2009sx} the authors consider in addition $r_{C_2}(E)\rightarrow C_2$ in order to remove the IR contribution to the entropy production. In our case, this has already been taken into account by the confining IR geometry.}. When the energy dependence of $x_{\perp}$ is of the form $x_{\perp}^2 =\sum_i c_i(E)c'_i(r)$, as it is usually the case, the last equation reduces to
\begin{align} \label{Sex2}
 S_{trap}(E) \sim x^2_{\perp}(E \rightarrow \infty, r=C_1).
\end{align}
In particular, the discussion above and equation (\ref{Eq}) imply that the geometries (\ref{bpl}) with a UV (constant) cut-off yield
\begin{subequations} \label{Ss}
 \be
S_{trap} \sim q_C \sim E^{\frac{2}{1+3a}}\sim s^{\frac{1}{1+3a}} \hspace{0.15in}a>1/3,\label{Ssa+}
\ee
\be
S_{trap} \sim q_C \sim E^{-\frac{2}{3(a-1)}}\sim s^{-\frac{1}{3(a-1)}} \hspace{0.15in}a\leq-1   \label{Ssa-}.
\ee
\end{subequations}
We remark that for $a>1$, the UV cut-off is placed on the upper bound of the integral in (\ref{Sex1}) as the boundary theory is at $r=\infty$ in this case. Evidently, this procedure, modifies the (center of mass) energy dependence (that we denote by $s$) of $S$. In the case of \cite{Gubser:2009sx},  it reduces to $S\sim s^{\frac{1}{6}}$ from $S\sim s^{\frac{1}{3}}$ (see equation (\ref{Eq}) and second line of table \ref{T2} for $a=-1$).
%%%%%%%%%%%%%%%%%%%%%%%
\subsection{Energy-dependent cut-off and the saturation scale}
%%%%%%%%%%%%%%%%%%%%%%%
We will now consider the shocks with uniform transverse dependence for simplicity.

In all examples we have analyzed, in the high-energy limit the trapped surface produces the following entropy
\begin{align} \label{cj1}
S_{trap} \sim \int_{r_C(E)} b_0^3(r) dr  \hspace{0.2in} s\rightarrow \infty
\end{align}
where $b_0(r)$ is the asymptotic form of $b$ for small $r$. $r_C=r_C(E)$ is the (lower) boundary of the trapped surface and it is determined by the boundary conditions (see e.g. (\ref{BCk3})). It is generically energy-dependent. We believe, that this behavior is much more general.

\begin{figure}[h!]
\centering
\includegraphics*[scale=0.7,angle=0,clip=true]{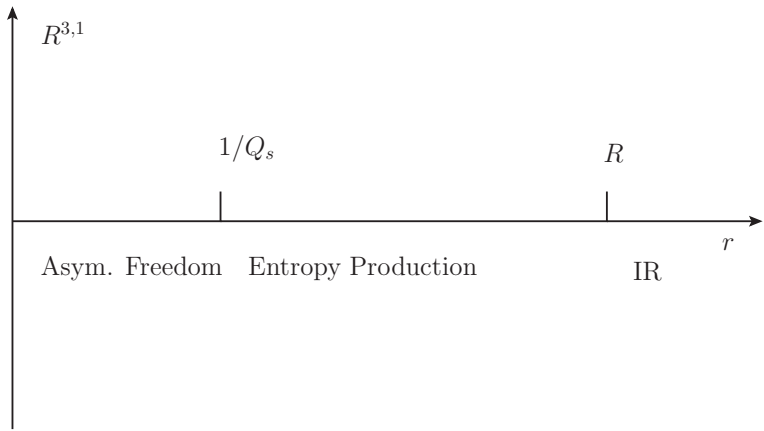} \hspace{1cm}
\caption{The entropy production occurs between the scales $\Lambda_{QCD} \sim 1/R$ and $Q_s$ which in the geometric language corresponds to $1/Q_s<r<R$. The $\Lambda_{QCD}$ has already been incorporated by the usage of a confining scale factor $b(r/R)$ as in ({\protect \ref{e^r2}}). The weak coupling regime on the other hand can not be implemented  rigorously in the framework of the AdS/CFT duality because string theory is not weakly coupled at high energies where 't Hooft coupling is small. The phenomenological $Q_s$ cut-off is proposed as a first approximation.}

\label{EnSc}
\end{figure}

Equation (\ref{cj1}) implies that most of the entropy originates in the UV part of the trapped surface. As the energy becomes larger this part enters into the region of the asymptotic freedom where the coupling is small and where we do not expect a large multiplicity to be produced. Therefore, as we have argued in the beginning of this section, we will impose asymptotic freedom by cutting-off the trapped surface at some $r_0$ as in \cite{Gubser:2008pc}. It is natural to expect that $r_0$ may be energy-dependent. We propose as a natural cut-off the saturation scale $Q_s$ (see figure \ref{EnSc}) by identifying $r_0 \sim 1/Q_s$ i.e.
%%%%%%%%%%%%%%%%%%%%%%%%%%%%%%%%%%%%%%%%%%%%%%%%
\begin{align} \label{cj2}
S_{trap}/\mbox{trans. area} \sim \int_{1/Q_s} b_0^3(r) dr  \mbox{    and    } S \sim \int_{1/Q_s} b^3(r)x^2_{\perp}(s \rightarrow \infty,r) dr,      \hspace{0.2in} s \rightarrow \infty.
\end{align}
In (\ref{cj2}) the scale $Q_s$ can be computed perturbatively and may be modeled by
\begin{align} \label{Qs}
Q_s^2(E) \approx (0.2 GeV)^2\times A^{1/3} (\sqrt{s_{NN}})^{2\lambda}
\end{align}
where $\sqrt{s_{NN}}$ is measured in GeV and denotes the c.m. energy in nucleon-nucleon collisions. In this equation $A$ is the atomic number while typically $\lambda$ lies in $[0.1,0.15]$ for energies at RHIC (and LHC) depending from the nature of the nuclear matter (pp, AA etc) participating in the collision \cite{Kovchegov:2010pw,JalilianMarian:2005jf,Albacete:2007yr,Mueller:2002zm,Triantafyllopoulos:2002nz,Lappi:2011gu,Albacete:2009fh,Levin:2011hr,Gelis:2010nm,Lublinsky:2011cw,Kowalski:2008sa,Dumitru:2001jn,Albacete:2010bs} and $\sqrt{s_{NN}}$ is measured in GeV's. The interval for the values of $\lambda$ is obtained from fitting data from independent processes, e.g Deep Inelastic Scattering (DIS) while they are close to the results predicted by analytical calculations \cite{Triantafyllopoulos:2002nz}.
{\it In the uniform transverse dependence case} and assuming that at small $r$ the geometry maps to $AdS_5$ then cutting the surface at $1/Q_s$ yields
\begin{align} \label{SQs}
S_{trap} \sim Q_s^2   \hspace{0.2in} s\rightarrow \infty.
\end{align}
It is interesting to note the $Q_s^2$ dependence for any scale factor $b(r)$ that has the asymptotic behavior $b_0=L/r$! The same dependence of entropy with $Q_s^2$ was found recently in \cite{Kutak:2011rb} for uniform transverse nuclei in the context of a different approach. In addition, multiplicity densities $dN/d\eta$ are proportional to $\sim Q_s^2$ as these are predicted on theoretical grounds from perturbative methods \cite{ Triantafyllopoulos:2002nz}.  $\eta$ denotes the pseudorapidity  while $dN/d\eta\sim Q^2_s$ describes the data quite well \cite{Lappi:2011gu,Levin:2011hr,Gelis:2010nm}.

However, when non-uniform transverse dependence is present, we can not reach such a general conclusion as the one provided by (\ref{SQs}) (compare with (\ref{cfS}) for instance).
\vspace{-0.1in}
\section{Results}\label{RS}
%%%%%%%%%%%%%%%%%%%%%%%%%%%%%%%%%%%%%%%%%%%%%%%%%%%%%%%%%%%%%%%%%%%%%%%%%%%%%%%%%%%%%%%
 %%%%%%%%%%%%%%%%%%%%%%%%%%%%%%%%%%%
\subsection{Behavior at Large $s$}\label{Ls}
\vspace{-0.1in}
%%%%%%%%%%%%%%%%%%%%%%%%%%%%%%%%%%%
Using the asymptotic form of the boundary conditions, equations (\ref{Eq}), (\ref{cond1}), and equations (\ref{ent1}), (\ref{Skfull}), (\ref{Sex1}), (\ref{Sex2}) and (\ref{cj2}) wherever appropriate we analytically or numerically compute the entropy production for all the cases we have considered. The results are summarized in tables \ref{T1}, \ref{T2} and \ref{T3}.

We remark that:
 \begin{itemize}
 \item (a) It makes sense to cut the surface at $r_s \sim  1/Q_s$ if and only if $r_s>r_{C_1}$ placing restrictions on the allowed values of $a$ in the power-low case (see caption of table \ref{T2} and footnote 16).

     \item (b) The power-law shocks for $a=-1$ (see \cite{Gubser:2008pc}) when the surface is cut at $1/Q_s$ yields $S_{trap}\sim r'Q_s  (r'A)^{1/3}s^{1/6}\sim A^{17/18}s^{0.24}$ (for AA; see (\ref{cfS})) where $r'$ is the radius of the nucleus A. This energy dependence describes data well (see plots of figure \ref{P1}). In addition, an almost linear dependence with the number of participants (see figure \ref{AL}), in PbPb collisions at $2.76$ TeV  has been observed \cite{Collaboration:2011rta}. The result for $2.76$ TeV concerns the ALICE experiment at LHC and is, up to this time, a preliminary result (see figure \ref{AL}).

         \item (c) The case of (\ref{e^r2}) has been analyzed for the exponential (in the transverse direction) profile corresponding to $k_1$. The results found fit satisfactorily  the RHIC data\footnote{See next subsection under which circumstances the fitting of (any) data is achieved.} up to $200$ GeV (see plots of figures  \ref{P3}).

              \item (d) The geometry corresponding to (\ref{bex}) is also interesting. In this case a trapped surface may be formed when a cut-off is placed in the UV region. The resulting entropy then behaves as $S_{trap}\sim \log^2(s)$ (see \cite{Kang:2004jd,Giddings:2002cd}) at large $s$ even when the cut-off of the trapped surface is energy-dependent. We remind the reader that the geometry (\ref{bex}) is a ``marginal" case corresponding to continuous spectra with a mass gap, \cite{ihqcd}.

                 \item(e) We have numerically derived a set of inequalities about the entropy production. In all the cases analyzed and compared (see below), the energies are taken identical and large\footnote{We do not assume a (UV) cut on the surface(s) in deriving the inequalities.}:

\begin{enumerate}
               \item For an exponential transverse profile corresponding to the geometry of (\ref{e^r2}) we find $S^{k_1}_{trap}>S_{trap}^{k_2}>S_{trap}^{k_3}>...$; $E R=$ fixed. Taking into account that $k_n$ sets the transverse size, we conclude that more dilute transverse distributions at fixed energy, result in more entropy.

               \item For an exponential transverse profile corresponding to the confining geometry of (\ref{e^r2}) we find that as $ER\to \infty$ with $E/k$ fixed, the entropy increases and becomes that of  the geometry $b=L/r$ (AdS$_5)$ (see (\ref{ineq})).

                  \item For the geometry $b=L/r$ we compare the (trapped) entropy of a  shock with exponential (transverse) profile (exp shock) with a shock having a power-like transverse profile, \cite{Gubser:2008pc} (GPY shock). We assume that both of the transverse profiles fall-off for $x_{\perp}>1/k$; the first falls-off exponentially (as $K_0(k x_{\perp})$) while the second as a power,  $1/(x^2_{\perp}+1/k^2)^3$. We find $S^{GYP}_{trap}>S^{exp}_{trap}$ and we conclude (again) that more dilute energy distributions produce more entropy at the same total energy.

               \item For uniform transverse distributions, the analysis of the geometries $b=L/r$ (AdS$_5)$ and (\ref{e^r2}) results in  $S^{AdS_5}_{trap}>S_{trap}^{L/r e^{-r2/R^2}}$. We conclude that confined matter produces less entropy than conformal matter at infinite coupling.
                    This is accord with basic intuition.
               We point out however, that the difference in the two entropies is subleading at high energy because most of the entropy is produced at the UV where the two geometries coincide ($L/r e^{-r^2/R^2}\approx L/r$ for small $r$.).

\end{enumerate}
             \end{itemize}

%%%%%%%%%%%%%%%%%%%%%%%%%%%%%%%%%%%%%%%%%%%%%%%%%%%%%%%%%%%%%%%%%%%%%%%%%%%%%%%%%%%%
\begin{table}[h]
\caption{Classification of trapped surfaces and entropy production $S=S_{trap}$ at high-energies of shocks with uniform transverse distribution. In all the cases, the last column assumes that $b(r)$ reduces to $L/r$ at small $r$. The first column of the table displays the large $r$ asymptotics. { The cut-off at the last two columns refers to a cut-off of the UV region of the trapped surface.}}
\vspace{0.1in}
\centering
\begin{tabular}{c|ccccc}
\hline\hline

b(r) &Confining &Sources & S with & S with &S with \\
 &  &  & no cut & cut at &cut at \\
  &  &  & &const.&$1/Q_s$ \\

\hline\hline
$e^{-(r/R)^{a}}$&  For $a>0$  &No  &  $ \sim s^{1/2}\log(s)^{\frac{1+a}{a}}$ & $\sim s^{1/2}\log(s)^{\frac{1+a}{a}}$ &$\sim Q_s^2$ \\
$e^{-(\frac{R}{r_0-r})^{a}}$&  For $a>0$  &No  &  $ \sim s^{1/2}\log(s)^{\frac{1-a}{a}}$ & $\sim s^{1/2}\log(s)^{\frac{1-a}{a}}$ &$\sim Q_s^2$ \\
$(r_0-r)^{a}$&  For $a >\frac{1}{3}$  &No  &  $ \sim s^{\frac{3a+1}{6a}}$ & $ \sim s^{\frac{3a+1}{6a}}$  &$\sim Q_s^2$ \\
$r^{a}$&  For $a\leq -1$  &No  &  $ \sim s^{\frac{3a+1}{6a}}$ & $ \sim s^{\frac{3a+1}{6a}}$ &$\sim Q_s^2$ \\
 \hline\hline
\end{tabular}
\label{T1}
\end{table}
%%%%%%%%%%%%%%%%%%%%%%%%%%%%%%%%%%%%%%%%%%%%%%
\vspace{-0.2in}
\begin{table}[h]
\caption{Classification of trapped surfaces  and entropy production at high-energies of shocks with non-trivial transverse dependence.
It is assumed that $Q_s^2 \sim s^{\lambda}$ with $\lambda=0.15$ (for AA collisions). }
\vspace{0.1in}
\centering
\begin{tabular}{c|ccccc}
\hline\hline
b(r) &Con- &Sources & S with & S with &S with \\
 & fining &  & no cut & cut at a &cut at \\
  &  &  & &const.&$1/Q_s$ \\
\hline\hline
$(r-r_0)^{a}$ \hspace{0.02in} $a>1/3$ &Yes&Yes & $\sim s^{\frac{3(a+1)}{2(3a+2)}}$& $\sim s^{\frac{1}{3a+1}}$  &    \\
$(r-r_0)^{a}$ \hspace{0.02in} $a \leq -1$&No&Yes & $\sim s^{\frac{(3a+1)}{6 a}}$& $\sim s^{\frac{1}{3(1-a)}}$  & $\sim s^{ \frac{ 2+3 \lambda (3a^2-1)}{6(1-a)}}$ \\
$ e^{-r/R}$ &Yes& Yes& $\sim s ^{1.66}\log^{1.17}(s)$& $\sim \log^2(s)$&$\sim \log^2(s)$\\ \\
 \hline\hline
\end{tabular}
\label{T2}
\end{table}

%%%%%%%%%%%%%%%%%%%%%%%%%%%%%%%%%%%%%%%%%%%%%%%
\begin{table}
\caption{Classification of trapped surfaces  and entropy production at high-energies of shocks with non-trivial transverse dependence. The entries of the first line-last three columns is a pure guess motivated from the results of $b\sim1/r$.
The second line corresponds to the normalizable $k_1$ mode.
The quantities $m(s)$ and $n(s)$
are slow functions of $s$ ranging (approximately) in the intervals [0.42,0.52] and [0.5.,1]
respectively as $\sqrt{s}$ increases in $[20,200]$  GeV.
In the same energy interval, $c_1$ ranges in $[300,775]$. An accurate fitting is found (see left plot of figure ({\protect \ref{P3}})).}
\centering
\begin{tabular}{c|ccccc}
\hline\hline
b(r) &Confining &Sources & S with & S with &S with \\
 &  &  & no cut & cut at a &cut at \\
  &  &  & &const.&$1/Q_s$ \\
\hline\hline
$\frac{L}{r}e^{-r/R}$ &Yes& Yes&$\sim s^{1/3} \log^2(s)$& $\sim s^{1/6} \log^2(s)$&$\sim s^{1/6} Q_s  \log^2(s)$\\
 \\
 $\frac{L}{r} e^{-r^2/R^2}$  &Yes& No &$ \sim( \sqrt{s})^{m(s)}\times$ &Not
 &Not
 \\
$\left( k_1=\frac{6}{R}\right)$&& &$  \log^{n(s)}(c_1\sqrt{s})$ &  Interesting
&Interesting
 \\ \\
%
% \vspace{0.1in}
 %
 & & & (Numerically)&
 &
 \\
 \hline\hline
\end{tabular}
\label{T3}
\end{table}
%%%%%%%%%%%%%%%%%%%%%%%%%%%%%%%%%%%%%%%%%%%%%%%%%%%%%%%%%%%%%%%%%%%%%%%%%%%%%
%%%%%%%%%%%%%%%%%%%%%%%%
\subsection{Fitting Data}
%%%%%%%%%%%%%%%%%%%%%%%%
\subsubsection{Relating $S_{trap}$ with multiplicities}\label{par}
%%%%%%%%%%%%%%%%%%%%%%%
The trapped surface analysis does not give the produced entropy but it  provides a lower bound
\begin{align} \label{bound}
S_{trap}\leq S_{prod.}.
\end{align}
Moreover there are several simplifying assumptions that remain between any comparison of the calculations done here and experimental data. We have spelled them out in the introduction and commented on how much each of them is expected to affect the connection with the data. In particular we have argued that for collisions and bulk geometries where the majority of the trapped surface area comes from the UV, AdS part of the geometry, we should still expect that the relative factor relating the final entropy release to the area of the trapped surface to be almost energy-independent.

We must also quantify the relation of the total multiplicity and the produced entropy. The total entropy is given by the  the number of charged particles $N_{ch}$ times $\sim 3/2$ to account for the neutral particles multiplied by $\sim 5$,  that is the entropy per particle \cite{Gubser:2008pc,Ochs:1996yf,Muller:2011ra}. Hence,
\begin{align} \label{SN}
S_{prod.} \approx 7.5 N_{ch}.
\end{align}
We have analyzed the two cases: (a) The case of the AdS geometry with a cut-off at the UV (at $\sim 1/Q_s$ where $Q_s$ the saturations sale). (b) The case of the confining IHQCD-like geometry (\ref{e^r2}) for $n=1$ (first excitation) without any UV cut-off.

It should be stressed here that a single parameter is used in these fits, namely the overall constant coefficient of the leading (large) s-dependence.

\subsubsection{Multiplicities for the AdS-$Q_s$ setup }\label{UVQs}
%%%%%%%%%%%%%%%%%%%%%%%%%%%%%%%%%i

This setup has an AdS metric, a GPY-like transverse profile and a UV cutoff at $r \sim 1/Q_s$ with $Q_s$ given by (\ref{Qs}).
The gravity parameters are chosen according to \cite{Gubser:2008pc} as $L^3/G_5\approx 1.9$. We need in addition the following relations
\begin{align} \label{md1}
1=0.197 \hspace{0.05in}\mbox{GeV.fm}  \hspace{0.2in}E=A\frac{\sqrt{s_{NN}}}{2}=A\frac{\sqrt{s}}{2}
\end{align}
where $A$ is the atomic weight of the participating nuclei in the collision and $\sqrt{s_{NN}}$ the center of mass energy/nucleon \footnote{We drop the subscript NN from $\sqrt{s_{NN}}$ from now on for simplicity.}.

%%%%%%%%%%%%%%%%%%%%%%%%%%%%%%%%%%%%%%%%%%%%%%%%
\begin{figure}[!th]
\centering
\includegraphics*[scale=0.72,angle=0,clip=true]{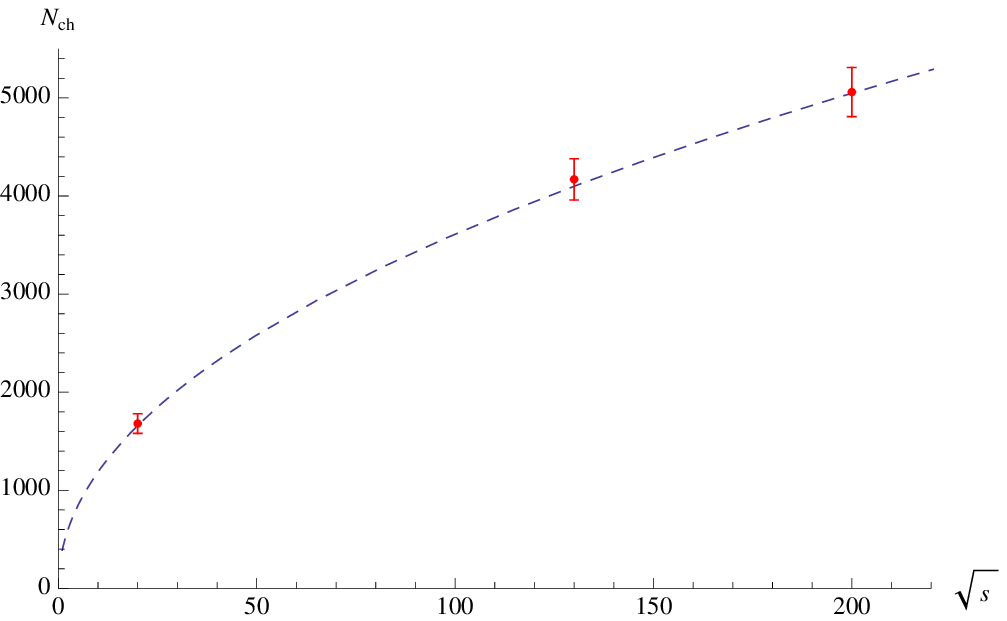} \hspace{0.1cm}
\includegraphics*[scale=0.72,angle=0,clip=true]{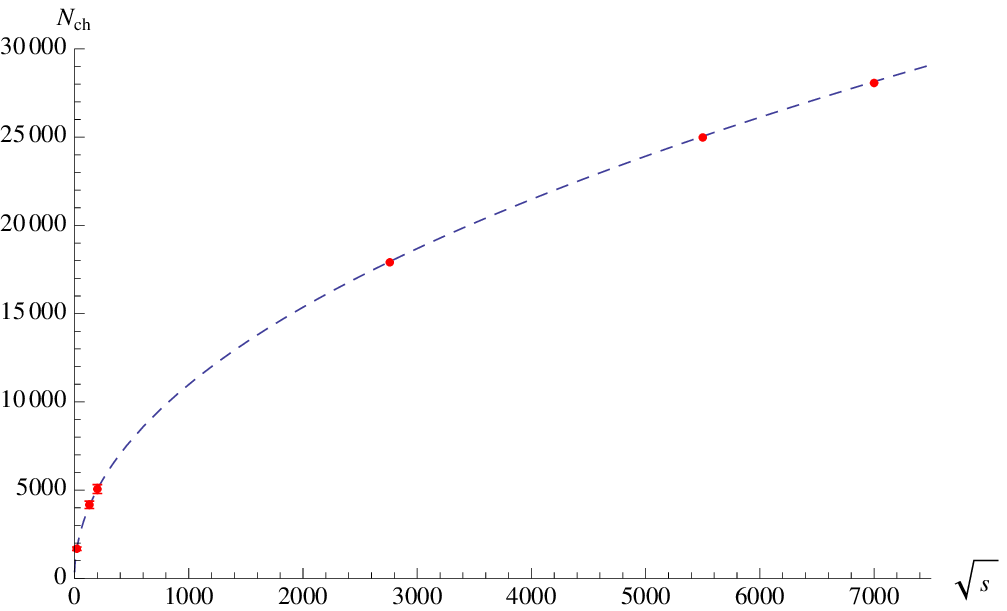}
\caption{Total multiplicities as functions of the c.m. energy measured in GeV.
 The red dots are experimental data from RHIC (AuAu collisions) taken from {\protect \cite{Back:2002wb} }
 with the error bars included while the theoretical curves are drawn using dashed lines. The same applies for figure {\protect \ref{P3}}.
 The left plot concerns the $AdS$ background as in {\protect \cite{Gubser:2008pc} }
 with a cut of the surface at the UV at $c/Q_s$ where $c \sim 1$ and transverse size $r'=r_{Au}\approx 8$ fm (with A$=197$; see ({\protect \ref{cfS}})).
 The right plot is the extension of the left plot at higher values of the energy for PbPb collisions (that is for A$=207$; see ({\protect \ref{cfS}})) with the points for $\sqrt{s}=2.76$, $5.5$ and $7$ TeV inserted.}
\label{P1}
\end{figure}
%%%%%%%%%%%%%%%%%%%%%%%%%%%%%%%%%%%%%%%%%%%%%%%%
We will cut-off the trapped surface at $r_s=c/Q_s$,  where $c$ a positive constant\footnote{As mentioned earlier, it makes sense to cut at $c/Q_s$ iff $c/Q_s>r_{C_1}=r'/2 \left(A\sqrt {s}G_5 r' /L^3 \right)^{-1/3}$ ($r'$ is the transverse nuclear size). For $c \sim 1$, A any (reasonable) value (see (\ref{Qs})), $r' \approx  (A/A_{Au})^{1/3} \times 8$ fm and $\sqrt{s}\geq 20$ measured in GeV, this condition is satisfied.}. One then computes (at large s)
\begin{align} \label{cfS}
S_{trap}&=2\times \frac{L^3}{4G_5}2\pi \int_{r_{C_1}\rightarrow c/Q_s}^{r_{C_2}} \frac{1}{r^3} \frac{x^2_{\perp}(r,Er')}{2}  \hspace{0.15in} \mbox{where} \hspace{0.15in} Er'=4\frac{L^3}{G_5}\left( \frac{x_{\perp}^2+(r-r')^2}{4 r r'}\right)^3                  \notag\\&
=\frac{\pi}{c} \left( \frac{L^3}{G_5}\right)^{2/3}(Q_sr' )(A r'\sqrt{s})^{1/3}
\approx \frac{1900}{c} \left (\frac{A}{A_{Au}}\right)^{17/18} \left(\frac{\sqrt{s}}{1\hspace{0.03in} GeV}\right)^{0.483}\Big|_{r_{Au}\approx 8 \hspace{0.02in} fm}
\end{align}
where we normalized the formula of $S_{trap}$ with $A_{Au}=197$ for AuAu collisions.  In normalizing, we used the fact that $r'$ is the transverse size of the colliding glueball\footnote{It corresponds to the position of the point-like source in the fifth dimension (see discussion above (\ref{ph})).} (beam) assuming that it satisfies the empirical law $(r'/r_{Au})=A^{1/3}/A_{Au}^{1/3}$ which applies for nuclei. In the last equality, we have used (\ref{Qs}) with $\lambda=0.15$ for AA collisions, (\ref{md1}) and $L^3/G_5=1.9$ \cite{Gubser:2008pc}.

The limits of the $r$ integration are found by setting $x_{\perp}=0$ (see figure \ref{TSB}). 
 The excellent fitting of plot \ref{P1} with RHIC data is achieved for A=A$_{Au}=197$ and $(1/c)\times$(overall coefficient\footnote{The meaning of the overall coefficient is discussed in subsection \ref{par}.})$\approx 1.54$ taking $r_{Au}=8$ fm\footnote{A different value for $r_{Au}$ but close to $8$ fm may still fit RHIC data.}. It is pleasing that both, the overall coefficient and $c$ can simultaneously be of order one. The extrapolation to higher energies is done in section \ref{pred} and the right plot of figure \ref{P1}.

%%%%%%%%%%%%%%%%%%%%%%%%%%%%%%%%
\subsubsection{Multiplicities for the IHQCD setup}\label{smM}
%%%%%%%%%%%%%%%%%%%%%%%%%%%%%%%%%%%%

This setup involves  a metric $b=L/r e^{-r^2/R^2}$ without a UV cut-off and an exponential transverse profile.
In this case, $S_{trap}$ is given by (\ref{Skfull}) for $n=1$ and has to be solved numerically.
The parameters of gravity  were determined  by matching lattice data in \cite{data} and we will use these in the sequel. Choosing $N_c=3$ we find

\begin{align} \label{md2}
k_1=\frac{6}{R}=3.1\hspace{0.05in}\mbox{GeV}  \hspace{0.2in}\frac{L^3}{\kappa_5^2}\approx 1.96,
\end{align}
where $k_1=m_1$ is the mass of the lightest spin-two glueball (second excitation).

%%%%%%%%%%%%%%%%%%%%%%%%%%%%%%%%%%%%%%%%%%%%%%%%
\begin{figure}[!th]
\centering
\includegraphics*[scale=0.72,angle=0,clip=true]{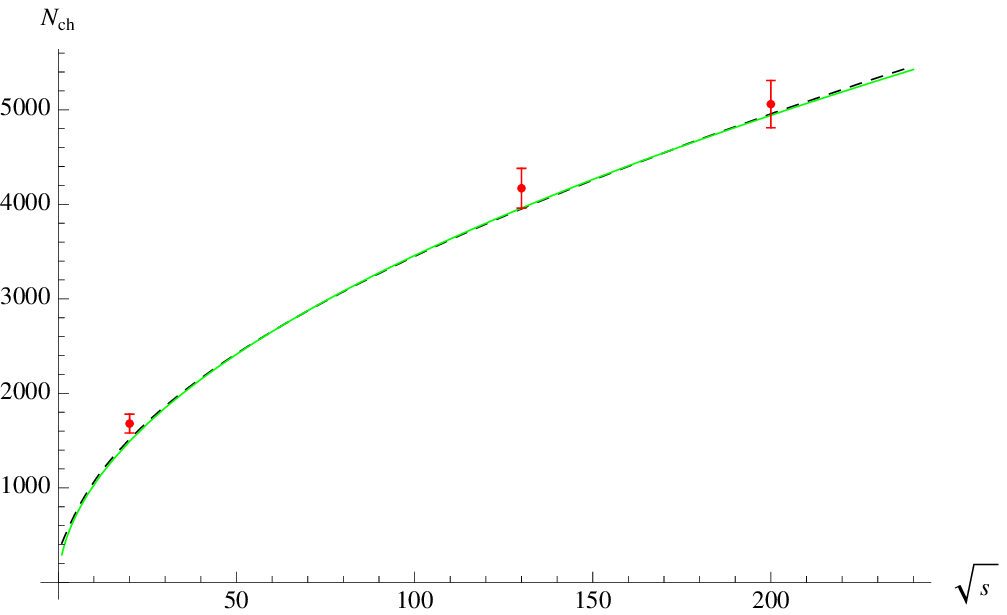} \hspace{0.1cm}
\includegraphics*[scale=0.72,angle=0,clip=true]{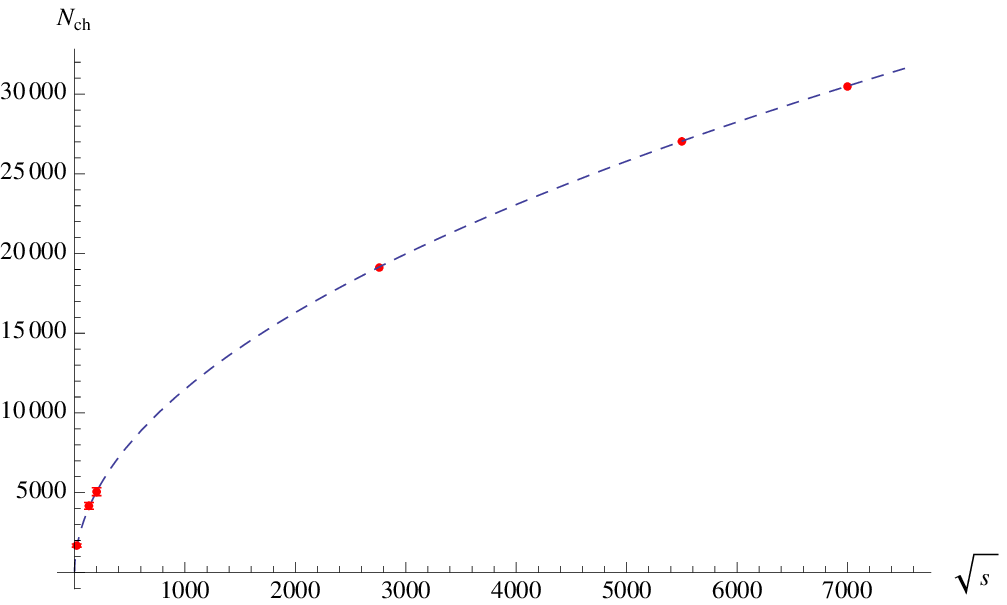} \hspace{0.1cm}
\caption{Both figures concern the numerical analysis of the surface (resulted from the first excitation) of ({\protect \ref{Skfull}})
for lower and higher energies (dashed plots). There is not a UV cut-off in this case. In the left figure, the green plot is given by ({\protect \ref{fitc}})
and exhibits the accurate approximation of the numerical plot found for A=A$_{Au}$. The overall coefficient of the numerical plot
has been chosen in order to fit the RHIC data. The right plot is the extension of the approximating plot at higher energies for A$=A_{Pb}$ and hence, according to ({\protect \ref{fitc}}), it is given by $N_{ch}=79.8\left(\sqrt{s} \right)^{0.451} \log^{0.718}\left(562  \sqrt{s}\right)$. The points for $\sqrt{s}=2.76$, $5.5$ and $7$ TeV are inserted.}
\label{P3}
\end{figure}
%%%%%%%%%%%%%%%%%%%%%%%%%%%%%%%%%%%%%%%%%%%%%%%%%%%%%%%%%%%%%%%%%%%%%%%%%%%%%%%

 The numerical result for multiplicities are plotted in figures \ref{P3} for weight A$=A_{Au}$ and compared with RHIC data.
 The left dashed plot is the result of our numerical analysis for A=A$_{Au}$ and energies up to $250$ GeV including the RHIC data. The agreement is satisfactory. The numerical result can be approximated very accurately (see green plot of figure \ref{P3}) by
\begin{align} \label{fitc}
N_{ch}=78.05 \left(\frac{A}{A_{Au}}~{\sqrt{s}\over 1~{\rm GeV}} \right)^{0.451} \log^{0.718}\left(534.9\frac{A}{A_{Au}} {\sqrt{s}\over 1~{\rm GeV}} \hspace{0.02in}\right)
\end{align}
where $\sqrt{s}$ is measured in GeV. In order to go higher in the energies, a more refined numerical analysis is needed. Hence, for the present work, we use the fitted curve given by equation (\ref{fitc}) in order to predict multiplicities  for higher energies. We do this in section \ref{pred} for the energies to be reached by LHC taking into account that for Pb A$=207$. The corresponding plot is the one on the right of figure \ref{P3}.

We remark that cutting the surface at some UV cut-off does not improve the fitting. In particular, for energies higher than $200$ GeV we either find very low multiplicities (for a constant cut-off) or very large multiplicities (for an energy dependent cut-off).

We close this section by noticing that in order to obtain a (more realistic) value for $ \frac{L^3}{\kappa_5^2}$ in the case of the trapped surface given by (\ref{Skfull}), the black hole ansatz for the scale factor of (\ref{e^r2}) should be solved.  We postpone  a detailed analysis with the precise IHQCD profiles for later work and choose the particular values mentioned below equation (\ref{SN}) in order to fit the data.

%%%%%%%%%%%%%%%%%%%%%%%%
\subsection{Multiplicities at LHC energies}\label{pred}
%%%%%%%%%%%%%%%%%%%%%%%%

In the previous section we have analyzed (a) the geometry of AdS$_5$ with a UV cut-off and (b) the first excited state of (\ref{e^r2}) without any UV cut-off. We have seen that both fit the RHIC data up to $200$ GeV satisfactorily (see plots of figures \ref{P1} and \ref{P3}). For the aforementioned geometries (a) and (b), we may extrapolate them at higher energy in order to assess what they predict for the multiplicities at the energies reached by LHC.

\begin{itemize}
 \item Geometry (a): Multiplying formula (\ref{cfS}) by $1.54$ (ignoring $c$; see subsection \ref{UVQs}), dividing over $7.5$ (see subsection \ref{par})  and taking $A=207$ for Pb central collisions we find $N^{Pb}_{ch}\approx18750$ for\footnote{All the energies that are mentioned in this subsection refer either to $\sqrt{s_{NN}} $ or  to $\sqrt{s_{pp}} $. } $2.76$ TeV, $N^{Pb}_{ch} \approx261800$ for $5.5$ TeV and $N^{Pb}_{ch} \approx29400$ for $7$ TeV.

For high-multiplicity\footnote{The notion of centrality in pp collisions is trickier to define. The best definition is to select high multiplicities in the final state. The definition of high multiplicity is ambiguous, but may still defined.}   proton-proton (pp) central collisions where A$=1$ we find $N^{p}_{ch}\approx 70$ for $0.9$ TeV, $N^{p}_{ch} \approx 110$ for $2.36$ TeV, $N^{p}_{ch} \approx 190$ for $7$ TeV and  $N^{p}_{ch} \approx 260$ for $14$ TeV. It is pointed out that extracting experimental results (for total $N_{ch}^p$) from ATLAS \cite{Collaboration:2011hd} and CMS \cite{Khachatryan:2010us,Khachatryan:2010nk} (for these energies) is not a trivial task. It involves model dependent procedures and Monte-Carlo simulations.

 \item Geometry  (b) The numerical analysis corresponding to (\ref{fitc}) (see right plot of \ref{P3}) for PbPb colisions predicts $N_{ch} \approx19100$ for $2.76$ TeV, $N_{ch}  \approx27000$ for $5.5$ TeV and $N_{ch} \approx 30500$ for $7$ TeV\footnote{A prediction for pp collisions using (\ref{fitc}) is less reliable for the moment because the errors induced from factors of $A_p/A_{Au}=1/197$ are larger than in PbPb collisions. To deal with this, a more refined numerical work is needed.}.

\end{itemize}

In the above results, we have  assumed exactly a zero impact parameter and hence these are zero centrality processes. This implies that we might predict slightly larger multiplicities than the upcoming data. The difference between the two predictions (at a given value of $\sqrt{s}$) for the two cases considered above is $\sim 6 \%$ for PbPb collisions. It is interesting that the first preliminary results from ALICE at LHC for $\sqrt{s}=$ 2.76 TeV and PbPb collisions give $N_{ch}\approx 17000\pm1000$ for $0-5\%$ centralities \cite{Collaboration:2011rta}. In order to make a better estimation, we consider the right plot of figure 7 in \cite{Collaboration:2011rta} which shows that the maximum number of participants for the particular measurements, is approximately $380$. Hence, plugging A$=380/2=190$ and $\sqrt{s}=2.76$ TeV in (\ref{cfS}) results $N_{ch}=17300$ which is extremely close to the (preliminary) measurements (see figure \ref{AL}).

%%%%%%%%%%%%%%%%%%%%%%%%%%%%%%%%%%%%%%%%%%%%%%%%%%%%%%%%%%%%%%%%%%%%%%%%%%%%%%%
\section{Outlook and discussion}\label{finally}
%%%%%%%%%%%%%%%%%%%%%%%%%%%%%%%%%%%%%%%%%%%%%%%%%%%%%%%%%%%%%%%%%%%%%%%%%%%%%%%

The initial motivation of this work was to estimate  the produced entropy of colliding shocks in different circumstances and obtain a qualitative answer for the results, in view of applications  to heavy-ion collisions. In particular, the goal was to compute $S_{trap}$ and hence (estimate) the particle multiplicities (see section \ref{par}). We have analyzed various dilaton-gravity geometries, with or without transverse dependence, with or without confinement and with or without UV cut-offs. The cut-offs were assumed energy dependent or energy independent. We conclude the following:

\begin{itemize}
\item Both our analytical calculations and our numerical analysis have shown that most of the entropy comes from the UV part of the geometry provided that the geometry is asymptotically AdS.

\item     There are geometries with a UV energy-independent cut-off that lead to an asymptotic $\sim \log^2(s)$ behavior for the entropy (see table \ref{T2}).

\item For uniform transverse distributions, the AdS$_5$ geometry, produces the least entropy among the geometries of section \ref{NoTD}. This result assumes that the non-conformal geometries in \ref{NoTD} survive until the boundary. They are therefore not asymptotically AdS.
     A similar result would be valid if the associated geometries are asymptotically AdS, but the energies such that the trapped surfaces do not penetrate the AdS region.  These results are summarized in table \ref{T1}.

\item We have constructed exact shock solutions, with non-trivial transverse distributions and point-like bulk sources, and have computed the entropy of the trapped surface (see table \ref{T2}). Inserting an energy-independent UV cut-off (see fifth column of table \ref{T2}), and choosing $a\approx 1$ %or $a\approx -1/3$ 
in the first %and second 
line \footnote{This case correspond to a power-like scale factor $b(r)\sim r^a$.},  results in $S\sim (\sqrt{ s})^{1/2}$ (see also (\ref{Ssa+})). The particular energy dependence seems to describe RHIC data as equations (\ref{cfS}), (\ref{fitc}) and figures \ref{P1} and \ref{P3} suggest. A similar power-law dependence  is obtained assuming a Landau hydrodynamical behavior \cite{Steinberg:2004vy} after the collisions . A characteristic of the geometries corresponding to $a=1$ and $-1/3$ is that they do not reduce to AdS$_5$ at the UV unlike the $a=-1$ geometry.

 \item  For geometries that have a mass gap and confinement, the entropy production, independent of transverse distribution, is subleading to that in AdS at the same total energy and transverse scale. Here it is assumed that all geometries are asymptotically AdS.

 \item The entropy production decreases  when higher glueballs collide (see \ref{ds}). This implies that more dilute transverse energy distributions  produce more entropy at fixed total energy ($ER=$ fixed, where $R$ can be identified with $1/ \Lambda_{QCD}$).

 \item We will denote the trapped surface areas as follows:  $S_{trap}^{GYP;AdS,k}$  for the GYP-like shockwave in (\ref{ph}) in AdS$_5$ where $k$ is the scale of the transverse profile,
 $S_{trap}^{k;Ads}$
 for an exponential transverse profile in AdS$_5$ as in (\ref{phik}),
 and $S_{trap}^{k_n}$ for an exponential transverse profile with scale $k_n$ for the confining metric (\ref{e^r2}). In all of the above the total energy, and the transverse scale are kept the same.

 We find the following inequalities for large $E$:
 \be \label{ineq}
 S_{trap}^{GYP;AdS}\gg S_{trap}^{k;Ads} \gg S_{trap}^{k_3}\gtrsim S_{trap}^{k_2} \gtrsim S_{trap}^{k_1} \Big|_{k=k_1=k_2=k_3} .
 \ee

 We conclude that the entropy production increases as the confinement scale ($\sim 1/R$ see subsection \ref{ds}) decreases provided that the transverse size (with respect to the energy) is kept fixed.

  \item  Equation  (\ref{cfS}) suggests that multiplicities should grow with $A$ as $A^{17/18}$. In fact, an almost linear dependence of $N_{ch}$ with the number of participants was observed in the recent ALICE experiments performed at $2.76$ TeV \cite{Collaboration:2011rta} for PbPb collisions. According to figure \ref{AL}, our result shows a similar behavior. The agreement becomes better as the number of participants increases, that is as the collision becomes more central which is the case that we assumed in this paper. We remark that the ALICE results are still preliminary.
 %%%%%%%%%%%%%%%%%%%%%%%%%%%%%%%%%%%%%%%%%%%%%%%%%%%%
 \begin{figure}[!th]
\centering
\includegraphics*[scale=0.85,angle=0,clip=true]{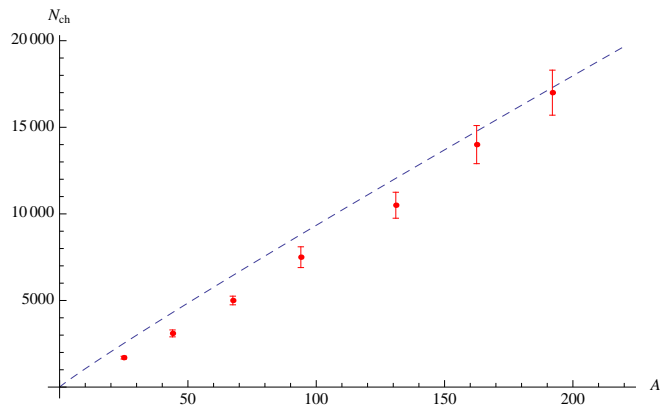} \hspace{-1.5cm}
\caption{The dashed plot refers to our theoretical prediction for PbPb collisions at $\sqrt{s}=2.76$ TeV.
It shows the total (charged) multiplicity $(N_{ch})$ as a function of A (see ({\protect \ref{cfS}})) at fixed energy $s$. The red points are data taken from reference {\protect \cite{Collaboration:2011rta}} with the error bars included: the horizontal coordinate of these points is taken to be $N_{part}/2$ where $N_{part}$ is the number of nucleons (of the two Pb nuclei) participating in the collision at the fixed value of  $\sqrt{s}=2.76$ TeV. The agreement of our theoretical prediction with the data seems to improve as $N_{part}$ increases, that is as the collision becomes more central.}
\label{AL}
\end{figure}
%%%%%%%%%%%%%%%%%%%%%%%%%%%%%%%%%%%%%%%%%%%%%%%%%%%%%
  \item We have derived two formulae for multiplicities, for the $AdS-Q_s$ setup in  (\ref{cfS}) and the IHQCD setup in  (\ref{fitc}).
         We normalize the multiplicities using RHIC data and we then compare with known LHC data. We find that in both cases they are very close to the $2.76$ TeV heavy-ion LHC data, \cite{Collaboration:2011rta} (see subsection \ref{pred}).

        \item An important puzzle of our analysis is that the scattering of the lowest lying $2^{++}$ glueballs does not seem to lead to a trapped surface.
\end{itemize}

For future work, we propose to use the shock of equation (\ref{promisefi}) and estimate the produced entropy. This shock has the advantage of localizing the bulk sources and in addition, it retains the ingredients of a mass gap and confinement. Also, more accurate calculations that will specify the gravity dimensionless parameter $L^3/G_5$ (see e.g. (\ref{md2})) could be made. Finally, it is important to understand the kind of black holes which give rise to the shocks (after boosting) and which in turn, mimic the nuclear matter in the colliders. This matter is best described by finite size black holes \cite{kt,daRocha:2006ei}. In particular, in \cite{kt} a description of the collisions in term of what is known about such black holes was described.
It would be interesting to make this picture precise, and in this numerical work will be required.

 \addcontentsline{toc}{section}{Acknowledgements}
\section*{Acknowledgements}\label{ACKNOWL}

We would like to thank P. Romatschke for participating in early stages of this work and for reading the manuscript, and J. Albacete,  W. Brooks, J. Butterworth, G. Dvali, Y. Foka, F. Gianotti,  J. Guimaraes da Costa, U. Heinz, Y. Konstantinou, Y. Kovchegov, A. Milov, A. Pilkinghton, P. Sphicas, A. Toia, D. Triantafyllopoulos and G. Veres for useful conversations and input.
A. T. would like to thank LPTENS for hospitality during the duration of this effort.
This work was partially supported by a European Union grant FP7-REGPOT-2008-1-CreteHEP
 Cosmo-228644, and PERG07-GA-2010-268246.

%%%%%%%%%%%%%%%%%%%%%%%%%%%%%%%%%%%%%%%%%%%%%%%%%%%%%%%%%%%%%%%%%%%%%%%%%%%%%%%%%%%
\newpage
\appendix
\renewcommand{\theequation}{A\arabic{equation}}
\setcounter{equation}{0}
\section{Introduction to the theory of trapped surfaces.}
\label{A}
%%%%%%%%%%%%%%%%%%%%%%%%%%%
The starting point is the shock-wave metric
\begin{align}\label{shgmn}
ds^2= b(r)^2 \left( -2dx^+dx^- + dx^2_{\perp}+dr^2+\phi_1(x^1,x^2,r)\delta(x^+)(dx^+)^2\right).
\end{align}
Associated with this shock-wave we parametrize the (half piece of the) trapped surface $S_1$ by
\begin{align}\label{1TS}
x^+=0  \hspace{0,2in} x^-+\frac{1}{2}\psi_1(x^1,x^2,r)=0
\end{align}
where $\psi_1$ will have to be determined. Before finding the differential equation satisfied by $\psi_1$ it seems necessary to perform the following coordinate transformation which eliminates the $\delta(x^+)$ from (\ref{shgmn}). In particular we use the transformation
\begin{align}\label{theta}
x^- \rightarrow x^-+\frac{1}{2}\phi_1(x^1,x^2,r)\theta(x^+)
\end{align}
which transforms (\ref{shgmn}) into
\begin{align}\label{shnew}
ds^2=  b(r)^2 \left( -2dx^+dx^- + dx^2_{\perp}+dr^2 -\theta(x^+) \sum_{i=1,2,r}( \nabla_i\phi_1dx^i)dx^+\right).
\end{align}
It is (sometimes) useful to exchange the functions $\phi_1$ and $\psi_1$ with $\Phi_1$ and $\Psi_1$ defined via
\begin{align}\label{1Psi}
\Phi_1=b(r)  \phi_1     \hspace{0,2in} \Psi_1= b(r) \psi_1.
\end{align}
The trapped surface associated with the shock $\phi_1$ can be found solving the equation
\begin{align}\label{TScond}
\theta=h^{\mu \nu} \nabla_{\mu} l^{(+1)}_{\nu}=0
\end{align}
where a few explanations are in order:

 \hspace{-0.3in}(a) The vector $l^{(+1)}_{\nu}$\footnote{The necessity of the presence of the superscript $+$ is explained in what follows.} can be generally expressed by
\begin{align}\label{lmu1}
 l^{(+1)}_{\mu} dx^{\mu}=Adx^+ +B(dx^- + \frac{1}{2}d\psi_1)=Adx^+ +B(dx^- + \frac{1}{2}(\sum_{i=1,2,r} \nabla_i\psi_1) dx^i)
\end{align}
which implies that
\begin{align}\label{lmu2}
 l^{(+1)}_{\mu} =\left( l^{(+1)}_+, l^{(+1)}_-, l^{(+1)}_1, l^{(+1)}_2, l^{(+1)}_r  \right)=
 \left( A,B,\frac{1}{2}B\nabla_1\psi_1,\frac{1}{2}B\nabla_2\psi_1,\frac{1}{2}B\nabla_r\psi_1\right)
 \end{align}
for some functions $A$ and $B$ which are determined by the following requirements: (i) $g^{\mu \nu} l^{(+1)}_{\mu}  l^{(+1)}_{\nu}=0$ where $g^{\mu \nu}$ is the (inverse) metric of (\ref{shgmn}), (ii) $ l^{(+1)t}>0$ and (iii) $ l^{(+1)-}<0$. The vector $ l^{(+1)}_{\mu} $ as defined is transverse to $S_1$, null, forward and outward. There is another vector, the  $ l^{(-1)}_{\mu} $ which is also transverse to $S_1$, null and forward but inward and can be taken to be
\begin{align}\label{l-}
 l^{(-1)}_{\mu} =\left( l^{(-1)}_+, l^{(-1)}_-, l^{(-1)}_1, l^{(-1)}_2, l^{(-1)}_r  \right)=\left(1,0,0,0,0\right).
 \end{align}
 (b) The covariant derivative is taken with respect to the metric (\ref{shnew}).

 \hspace{-0.33in}
(c) The tensor $h_{\mu \nu}$ is the projective metric to $l^{(+1)}_{\nu}$ and is given by
\begin{align}\label{hmn}
h_{\mu \nu}=g_{\mu \nu}-\frac{l^{(+1)}_{\mu}l^{(-1)}_{\nu}+l^{(+1)}_{\nu}l^{(-1)}_{\mu}}{g^{\rho \sigma}l^{(+1)}_{\rho}l^{(-1)}_{\sigma}}
\end{align}
where $g^{\mu \nu}$ is the inverse metric of equation (\ref{shnew}). This tensor clearly has the property
\begin{align}\label{pr}
 h^{\mu \nu}l^{(+1)}_{\mu}=0.
 \end{align}
We are now in position to follow the explanations (a), (b) and (c) mentioned above.

\underline{ Explanation (a):} From $g^{\mu \nu} l^{(1)}_{\mu}  l^{(1)}_{\nu}=0$ and the metric (\ref{shnew}) one finds that
\begin{align}\label{l=0}
B\left( A- B \frac{1}{8 } \sum_{i=1,2,r}\left[\nabla_i (\frac{ \Psi_1}{b(r)}-\frac{ \Phi_1}{b(r)})\times \nabla_i  (\frac{ \Psi_1}{b(r)}-\frac{ \Phi_1}{b(r)})\right] \right)=0.
\end{align}
 We emphasize that (i) the solution, $B=0$ does not satisfy the rest requirements of  $l^{(+1)}_{\nu}$ and hence it is discarded. (ii) Both of the subscripts $(_i)$ in the $\nabla_i$'s of (\ref{l=0}) are lower indices. We point out that condition (\ref{l=0}) does not specify $A$ and $B$ uniquely but it only specifies their ratio
\begin{align}\label{A/B}
 \frac{A}{B} =\frac{1}{8}  \sum_{i=1,2,r}\left[\nabla_i (\frac{ \Psi_1}{b(r)}-\frac{ \Phi_1}{b(r)})\times \nabla_i  (\frac{ \Psi_1}{b(r)}-\frac{ \Phi_1}{b(r)})\right].
 \end{align}
Hence a choice for $A$ and $B$ is
\begin{align}\label{AB}
A=-  \sum_{i=1,2,r}\left[\nabla_i (\frac{ \Psi_1}{b(r)}-\frac{ \Phi_1}{b(r)})\times \nabla_i  (\frac{ \Psi_1}{b(r)}-\frac{ \Phi_1}{b(r)})\right]           \hspace{0.2in} B=-8
\end{align}
which is proportional (by the factor $\frac{1}{b(r)^2}\rightarrow  \frac{r^2}{L^2}$) with (101) of \cite{Gubser:2008pc}\footnote{Up to a factor of 2 for $B$ which is traced to the authors different convention for the light-cone coordinates.}. One may wonder that this arbitrariness in choosing $A$ and $B$ may invalidate the procedure that determines the trapped surface and in particular equation (\ref{TScond}). We show later that this is not the case and that (\ref{TScond}) is invariant under rescalings of the vector $l^{(+1)}_{\mu} $. Now, according to (\ref{lmu2}), the choice (\ref{AB}) for $A$ and $B$ implies
\begin{align}\label{lmu3}
& l^{(+1)}_{\mu} =\left( l^{(+1)}_+, l^{(+1)}_-, l^{(+1)}_1, l^{(+1)}_2, l^{(+1)}_r  \right) \notag\\&
=-  \left( 8, \sum_{i=1,2,r}\left[\nabla_i(\frac{ \Psi_1}{b(r)}-\frac{ \Phi_1}{b(r)}) \nabla_i (\frac{ \Psi_1}{b(r)}-\frac{ \Phi_1}{b(r)})  \right]  ,4\nabla_1(\frac{\Psi_1}{b(r)}),4\nabla_2(\frac{\Psi_1}{b(r)}),4 \nabla_r(\frac{\Psi_1}{b(r)})\right).
 \end{align}

\underline{ Explanation (b):} One has then to compute the covariant derivative of $ l^{(+1)}_{\mu}$ with respect to the metric (\ref{shnew}) forming the tensor $\nabla_{\nu}  l^{(+1)}_{\mu} $. We do not display explicitly this tensor.

\underline{ Explanation (c):} The final step is to contract $\nabla_{\nu}  l^{(+1)}_{\mu} $ with $h^{\mu \nu}$. This  yields
\begin{align}\label{box1}
 (\Box_{AdS_3}-F(b(r))) (\Psi_1-\Phi_1)=0 \hspace{0.2in}F(b(r))= \frac{\partial_r(b(r) b'(r))}{b(r)^4}
\end{align}
where $\Box_{AdS_3}$ is defined with respect to the metric
\begin{align}\label{AdS3}
ds^2= b(r)^2  \left( dx^2_{\perp}+dr^2\right).
\end{align}

Associated with a second shock $\phi_2$ moving along $x^-$ there exists the surface $S_2$ constructed from a second vector $l_{\mu}^{(+2)}$, the analogue of $l_{\mu}^{(+1)}$. From symmetry considerations, $l_{\mu}^{(+2)}$ can be immediately found from $l_{\mu}^{(+1)}$ (see (\ref{lmu3})) and hence reads

\begin{align}
 l^{(+2)}_{\mu}
=-  \left(  \sum_{i=1,2,r}\left[\nabla_i(\frac{ \Psi_2}{b(r)}-\frac{ \Phi_2}{b(r)}) \nabla_i (\frac{ \Psi_2}{b(r)}-\frac{ \Phi_2}{b(r)}) \right ]  ,8,4\nabla_1(\frac{\Psi_2}{b(r)}),4\nabla_2(\frac{\Psi_2}{b(r)}),4 \nabla_r(\frac{\Psi_2}{b(r)})\right).
 \end{align}

\underline{ Boundary conditions on $C=S_1\cap S_2$:}

 \vspace{0.1in}

  \hspace{-0.33in} $\Psi_1|_C=\Psi_2|_C$ (continuity of $S$ on $C$) and $  \sum_{i=1,2,r}\left[\nabla_i(\frac{ \Psi_1}{b(r)}-\frac{ \Phi_1}{b(r)}) \nabla_i (\frac{ \Psi_1}{b(r)}-\frac{ \Phi_1}{b(r)}) \right ] |_C=8=  \sum_{i=1,2,r}\left[\nabla_i(\frac{ \Psi_2}{b(r)}-\frac{ \Phi_2}{b(r)}) \nabla_i (\frac{ \Psi_2}{b(r)}-\frac{ \Phi_2}{b(r)}) \right ]  |_{C}$ and $\nabla_i \Psi_1|_C=\nabla_i \Psi_2|_C, i=1,2,r$ (by requiring  $ l^{(+1)}_{\mu}|_{C}= l^{(+2)}_{\mu}|_{C}$). Assuming for simplicity identical shocks, we have $\Phi_1=\Phi_2 \equiv\Phi$ and by symmetry $\Psi_1=\Psi_2 \equiv \Psi=0$ on $C$. Hence the boundary conditions we get finally are
 \begin{align}\label{BC1}
\Psi|_C=0 \hspace{0.4in}  \sum_{i=1,2,r}\left[ \left (\frac{1}{b(r)}\nabla_i \Psi-\nabla_i( \frac{1}{b(r)} \Phi)\right) \left(\frac{1}{b(r)} \nabla_i \Psi-\nabla_i(\frac{1}{b(r)}  \Phi)\right)\right] |_{C}=8
 \end{align}
 Equation (\ref{BC1}) may be simplified. Taking also into account that in the pre-collision time ($x^{\pm}<0$)  $\Phi$ vanishes (see (\ref{theta}) for $x^+<0$), equation (\ref{BC1}) reduces to
 \begin{align}\label{BC}
\Psi|_C=0 \hspace{0.4in}  \frac{1}{b(r)^2} \sum_{i=1,2,r}\left[ \nabla_i \Psi \nabla_i \Psi \right]  |_{C}=8.
 \end{align}

Finally, we prove that (\ref{TScond}) is invariant under rescalings of $l^{(+1)}_{\mu}$ by an arbitrary (differentiable) function $\alpha(x^{\nu})$:
We consider equation (\ref{TScond}) with $l^{(+1)}_{\mu} \rightarrow \alpha(x^{\nu})l^{(+1)}_{\mu}$. We have $\theta=h^{\mu \nu} \nabla_{\mu} l^{(+1)}_{\nu} \rightarrow \theta'= h^{\mu \nu} \nabla_{\mu} (\alpha l^{(+1)}_{\nu})$. Using the Leibniz rule of the covariant derivative we get $\theta'=\alpha h^{\mu \nu} \nabla_{\mu}(l^{(+1)}_{\nu})+h^{\mu \nu} l^{(+1)}_{\nu} \partial_{\mu}(\alpha)$. Using the projective property of  $h^{\mu \nu}$ on $l^{(+1)}_{\mu}$, equation (\ref{pr}), we find that the second term of last equality vanishes yielding $\theta'=\alpha h^{\mu \nu} \nabla_{\mu}(l^{(+1)}_{\nu})=\alpha \theta=0$ since $\theta=0$ by assumption $\Box$

%%%%%%%%%%%%%%%%%%%%%%%%%%%%%%%%%%%%%%%%%%%%%%%%%%%%%%%%%%%%%%%%%%%%%%%%%%%%%%%%%%%%
%\newpage
%\appendix
\renewcommand{\theequation}{B\arabic{equation}}
\section{Localizing the bulk sources}
\label{B}
%%%%%%%%%%%%%%%%%%%%%%%%%%%
The shocks we have considered in subsection \ref{Ltd} correspond to a non-localized bulk source in the $r$ direction (see (\ref{nhom})). In order to localize this source we begin from (\ref{nhom}) and consider appropriate linear combination of solutions. This is not hard as one merely has to use the completeness relation of $g_k$ which  schematically has the form
\begin{align} \label{cr}
\int dk \hspace{0.02in}C(k) g_k(r)g_k(r')=\delta(r-r').
\end{align}
This would imply that the shock $\phi_k$ should be given by
\begin{align} \label{lf}
\phi=\int dk \hspace{0.02in}C(k) K_0(k x_{\perp})g_k(r)g_k(r').
\end{align}
In particular the completeness relation
\begin{align} \label{Lc}
\sum_{n=0}^{\infty} \frac {n!} {\Gamma(n+a+1)} x'^a e^{-x'} L_n^{(a)}(x')L_n^{(a})(x)=\delta(x-x')
\end{align}
for the Laguerre polynomials implies that the correctly normalized shock\footnote{Such that the gauge tensor $T_{++}$ integrates to $E$.} $\phi$ should be given by\footnote{For the AdS$_5$ geometry, the sources are localized choosing $C(k)=k J_2(kr')$. The shock then becomes that of \cite{Gubser:2008pc}. This can be shown using results of reference \cite{GR}.}
\begin{align} \label{promisefi}
\phi=\frac{6E\kappa^2_5 R^2}{\pi L^3}  & \left (3\frac{r^2}{R^2} \frac{r'^2}{R^2} \right)^2\delta(x^+) \notag\\&
\times \sum_{n=0}^{\infty}
\frac {n!} {(n+2)!}K_0\left( \frac{x_{\perp}}{R} \sqrt{12(n+2)}\right)L_n^{(2)}\left(3\frac{r'^2}{R^2}\right)L_n^{(2)}\left(3\frac{r^2}{R^2}\right).
\end{align}
In arriving to (\ref{promisefi}) we have used
\begin{align} \label{ids}
\int K_0\left( \frac{x_{\perp}}{R} \sqrt{12(n+2)}\right) d^2{\bf x_{\perp}}= \frac{2\pi R^2}{12(n+2)}, \hspace{0.15in}   L_n^{(2)}(0)=\frac  {(n+2)!} {n!2}\hspace{0.1in}, \notag\\
 \sum_{n=0}^{\infty}
\frac {1} {(n+a)}L_n^{(a)}\left(3\frac{r'^2}{R^2}\right)=\Gamma(a)  \left ( \frac{R^2}{3r'^2} \right)^a.
\end{align}
%%%%%%%%%%%%%%%%%%%%%%%%%%%%%%%%%%%%%%%%%%%%%%%%%%%%%%%%%%%%%%%%%%%%%%%%%%%%%%%%%%%%%
%\appendix
\renewcommand{\theequation}{C\arabic{equation}}
\section{Proof of formula ({\protect \ref{BCk3}}) }
\label{C}
%%%%%%%%%%%%%%%%%%%%%%%%
The idea is to note that the boundary of the surface should contain the point $(r_C,x_{\perp C}=0)$ because the source is located at $x_{\perp C}=0$ while both the terms $(\partial_r \psi_k)$ and $(\partial_{x_{\perp}} \psi_k)$ should be finite at $x_{\perp C}$. We also state the following relations
\begin{subequations}\label{Wk}
\be
(\partial_{x_{\perp}}K_0(k x_{\perp}))I_0(k x_{\perp})-K_0(k x_{\perp})  (\partial_{x_{\perp}}I_0(k x_{\perp}))=-\frac{1}{x_{\perp}}+(k'-k)O_{k'}(x_{\perp}),\label{Wkx}
 \ee
 \be
 \lim_ {x _{\perp C}  \to 0}  I_0(k x_{\perp})=1+O(x_{\perp}^2) \label{Io}.
\ee
\end{subequations}
Beginning from $(\partial_{x_{\perp}} \psi_k)$ we have that
\begin{align}\label{psix}
\lim_ {x _{\perp C}  \to 0}   \partial_{x_{\perp}} \psi_k = \frac{E\kappa_5^2}{L^3k^2} \lim_ {x _{\perp C}  \to 0} \left( \frac{g_1(k r_C)}{x _{\perp C} } + \sum_{k'} (k'-k)O_{k'}(x_{\perp C}) \right)
\end{align}
where we have used (\ref{Wk}) in order to simplify the last expression. It is evident that $g_1(k r_C)$ should have (at least) two distinct real roots in order to have a trapped surface (see figure \ref{TSB}). For reasonable shocks, the $g_1$ decays for small $r$ as $r^4$ and hence the one root is at $r_C=0$. By assumption of the claim (see subsection \ref{Ltd}) there exists one more real root call it $r_{C_2}$ and as a result the following is true
\begin{align} \label{g10}
\lim_{r_C  \to r_{C_2}} g_1(k r_C)=0
\end{align}

We now compute $\partial_r \psi_k$ again for $x_{\perp C} \rightarrow 0$ and hence at $r_c \rightarrow r_{C_2}$. We have
\begin{align} \label{psir}
\hspace{0.2in}\lim_{x_{\perp C} \to 0, r_C \to r_{C_2}}\partial_r(\psi_k)&=\frac{E\kappa_5^2}{L^3k^2}
\lim_{x_{\perp C} \to 0, r_C \to r_{C_2}}  \frac{ K_0(k x_{\perp}) } {   \sum_{k'} \left(C_{k'}^1g_1(k' r_C )+C_{k'}^2g_2(k' r_C) \right)I_0(k' x_{\perp C}) }  \notag\\&
\hspace{0in} \times \Bigg[ \sum_{k'}I_0(k' x_{\perp})    \big[  g_1'(k r_C)\left(C_{k'}^1g_1(k' r_C )+C_{k'}^2g_2(k' r_C) \right) \notag\\&
- g_1(k r_C) \left(C_{k'}^1g_1'(k' r_C )+C_{k'}^2g_2'(k' r_C) \right)\big] \Bigg]\notag\\&
 =\frac{E\kappa_5^2}{L^3k^2}\lim_{x_{\perp C} \to 0, r_C \to r_{C_2}}  K_0(k x_{\perp}) g_1'(k r_C)
\end{align}
where in the second equality we have used (\ref{g10}). The simplified expression for $\partial_r(\psi_k)$ implies a logarithmic divergence when $x_{\perp} \rightarrow 0$ because  $\lim_{r_C \to r_{C_2}}g_1'(k r_c) \neq 0$ by the hypothesis of the claim that there are no multiple roots for $g_1$\footnote {With the exception for $r=0$ which is a fourth-root.}. The fact that the right-hand side of (\ref{psir}) is independent on the arbitrary coefficients $C_k^1$, $C_k^2$ would yield to the naive conclusion that the whole second term of (\ref{psik}) should be absent. However, looking more carefully inside the sum of the numerator of (\ref{psir}), one realizes that the terms for $g_1(k x_{\perp C})=g_1(k' x_{\perp C})$ do not actually participate in the sum and hence they are allowed. Thus the trapped surface equation is given by
\begin{align} \label{psik2}
\psi_k&= \phi_k(r,x_{\perp })-\frac{g_1(k r) I_0(k x_{\perp})  } { g_1(k r_C)  I_0(k x_{\perp C})  }  \phi_k(r_C,x_{\perp C })\notag\\&
= \frac{E\kappa_5^2}{L^3k^2}g_1(k r)\left(  K_0(k x_{\perp }) -  \frac{ I_0(k x_{\perp }) }{ I_0(k x_{\perp C}) } K_0(k x_{\perp C}) \right)
\end{align}
yielding to the trapped surface boundary determined by the condition (\ref{BCk3}) completing the proof of the claim.

%%%%%%%%%%%%%%%%%%%%%%%%%%%%%%%%%%%%%%%%%%%%%%%%%%%%%%%%%%%%%%%%%%%%%%%%%%%%%%%%%%%%%


\newpage
\begin{thebibliography}{99}


  %\cite{Kiritsis:2011qv}
\bibitem{kt}
  E.~Kiritsis, A.~Taliotis,
{\em   ``Mini-Black-Hole production at RHIC and LHC,''}
  \hri{1110.5642}{[hep-ph]}.
  %%CITATION = APHUE,A24,51;%%



%\cite{Kang:2004jd}
\bibitem{Kang:2004jd}
  K.~Kang, H.~Nastase,
{\em   ``High energy QCD from Planckian scattering in AdS and the Froissart bound,''}
  Phys.\ Rev.\  {\bf D72}, 106003 (2005).
\hre{hep-th}{0410173}.

%\cite{Giddings:2002cd}
\bibitem{Giddings:2002cd}
  S.~B.~Giddings,
  {\em ``High energy QCD scattering, the shape of gravity on an IR brane, and  the
  Froissart bound,''}
  Phys.\ Rev.\  D {\bf 67}, 126001 (2003)
\hre{ hep-th}{0203004}.
  %%CITATION = PHRVA,D67,126001;%%

%\cite{Lin:2010cb}
\bibitem{Lin:2010cb}
  S.~Lin, E.~Shuryak,
  {\em ``On the critical condition in gravitational shock wave collision and heavy ion collisions,''}
  Phys.\ Rev.\  {\bf D83}, 045025 (2011).
  \hri{1011.1918}{[hep-th]}.

%\cite{Kovchegov:2007pq}
\bibitem{Kovchegov:2007pq}
  Y.~V.~Kovchegov, A.~Taliotis,
{\em   ``Early Time Dynamics in Heavy-Ion Collisions from AdS/CFT Correspondence,''}
  Phys.\ Rev.\  {\bf C76}, 014905 (2007).
 \hri{0705.1234}{[hep-ph]}.

%\cite{Spillane:2011yf}
\bibitem{Spillane:2011yf}
  M.~Spillane, A.~Stoffers, I.~Zahed,
 {\em  ``Jet quenching in shock waves,''}
  \hri{1110.5069}{[hep-th]}.

    %\cite{Sfetsos:1994xa}
\bibitem{Sfetsos:1994xa}
  K.~Sfetsos,
{\em   ``On gravitational shock waves in curved space-times,''}
  Nucl.\ Phys.\  B {\bf 436}, 721 (1995)
 \hre{hep-th}{9408169}.
  %%CITATION = NUPHA,B436,721;%%


%\cite{Albacete:2008vs}
\bibitem{Albacete:2008vs}
  J.~L.~Albacete, Y.~V.~Kovchegov, A.~Taliotis,
  {\em ``Modeling Heavy Ion Collisions in AdS/CFT,''}
  JHEP {\bf 0807}, 100 (2008).
 \hri{0805.2927}{[hep-th]}.


%\cite{Khlebnikov:2010yt}
\bibitem{Khlebnikov:2010yt}
  S.~Khlebnikov, M.~Kruczenski, G.~Michalogiorgakis,
  {\em ``Shock waves in strongly coupled plasmas,''}
  Phys.\ Rev.\  {\bf D82}, 125003 (2010).
  {\hri 1004.3803} {[hep-th]}.

  %\cite{Khlebnikov:2011ka}
\bibitem{Khlebnikov:2011ka}
  S.~Khlebnikov, M.~Kruczenski, G.~Michalogiorgakis,
  {\em ``Shock waves in strongly coupled plasmas II,''}
  JHEP {\bf 1107}, 097 (2011).
  \hri {1105.1355}{[hep-th]}.

%\cite{Chesler:2010bi}
\bibitem{Chesler}
  P.~M.~Chesler, L.~G.~Yaffe,
{\em   ``Holography and colliding gravitational shock waves in asymptotically AdS$_5$ spacetime,''}
  Phys.\ Rev.\ Lett.\  {\bf 106}, 021601 (2011).
 \hri{1011.3562} {[hep-th]}.

  %\cite{Heller:2011ju}
\bibitem{Heller:2011ju}
  M.~P.~Heller, R.~A.~Janik, P.~Witaszczyk,
{\em   ``The characteristics of thermalization of boost-invariant plasma from holography,''}
\hri{1103.3452}{[hep-th]}.

%\cite{Albacete:2009ji}
\bibitem{Albacete:2009ji}
  J.~L.~Albacete, Y.~V.~Kovchegov, A.~Taliotis,
{\em   ``Asymmetric Collision of Two Shock Waves in AdS(5),''}
  JHEP {\bf 0905}, 060 (2009).
 \hri{0902.3046}{[hep-th]}.


%\cite{Aref'eva:2009kw}
\bibitem{Aref'eva:2009kw}
  I.~Y.~.Aref'eva, A.~A.~Bagrov, L.~V.~Joukovskaya,
  {\em``Critical Trapped Surfaces Formation in the Collision of Ultrarelativistic Charges in (A)dS,''}
  JHEP {\bf 1003}, 002 (2010).
  \hri{0909.1294}{[hep-th]}.



%\cite{Aref'eva:2009wz}
\bibitem{Aref'eva:2009wz}
  I.~Y.~.Aref'eva, A.~A.~Bagrov, E.~A.~Guseva,
 {\em  ``Critical Formation of Trapped Surfaces in the Collision of Non-expanding Gravitational Shock Waves in de Sitter Space-Time,''}
  JHEP {\bf 0912}, 009 (2009).
  \hri{0905.1087}{ [hep-th]}.


  %\cite{Wu:2011yd}
\bibitem{Wu}
  B.~Wu, P.~Romatschke,
{\em   ``Shock wave collisions in AdS5: approximate numerical solutions,''}
\hri{1108.3715}{[hep-th]}.

%\cite{Albacete:2008ze}
\bibitem{Albacete:2008ze}
  J.~L.~Albacete, Y.~V.~Kovchegov, A.~Taliotis,
{\em   ``DIS on a Large Nucleus in AdS/CFT,''}
  JHEP {\bf 0807}, 074 (2008).
\hri{0806.1484}{[hep-th]}.


%\cite{Taliotis:2009ne}
\bibitem{Taliotis:2009ne}
  A.~Taliotis,
{\em   ``DIS from the AdS/CFT correspondence,''}
  Nucl.\ Phys.\  {\bf A830}, 299C-302C (2009).
\hri{0907.4204}{[hep-th]}.

  %\cite{Grumiller:2008va}
\bibitem{Grumiller:2008va}
  D.~Grumiller, P.~Romatschke,
{\em   ``On the collision of two shock waves in AdS(5),''}
  JHEP {\bf 0808}, 027 (2008).
\hri{0803.3226} {[hep-th]}.


%\cite{Taliotis:2010pi}
\bibitem{Taliotis:2010pi}
  A.~Taliotis,
{\em   ``Heavy Ion Collisions with Transverse Dynamics from Evolving AdS Geometries,''}
  JHEP {\bf 1009}, 102 (2010).
\hri{1004.3500}{[hep-th]}.


%%%%%%%%%%%%%%%%%%%%%%%%%%%%%%
%\cite{CasalderreySolana:2011us}
\bibitem{CasalderreySolana:2011us}
  J.~Casalderrey-Solana, H.~Liu, D.~Mateos, K.~Rajagopal, U.~A.~Wiedemann,
{\em  ``Gauge/String Duality, Hot QCD and Heavy Ion Collisions,''}
   \hri{1101.0618}{[hep-th]}.


 %\cite{Edelstein:2009iv}
\bibitem{Edelstein:2009iv}
  J.~D.~Edelstein, J.~P.~Shock, D.~Zoakos,
  {\em``The AdS/CFT Correspondence and Non-perturbative QCD,''}
  AIP Conf.\ Proc.\  {\bf 1116}, 265-284 (2009).
\hri{0901.2534}{[hep-ph]}.


 %\cite{Taliotis:2010kx}
\bibitem{Taliotis:2010kx}
  A.~Taliotis,
 {\em  ``$q{\bar q}$ Potential at Finite T and Weak Coupling in ${\cal N}=4$,''}
  Phys.\ Rev.\  {\bf C83}, 045204 (2011).
  \hri{1011.6618}{[hep-th]}.

  %\cite{Bernamonti:2011vm}
\bibitem{Bernamonti:2011vm}
  A.~Bernamonti, R.~Peschanski,
  {\em ``Time-dependent AdS/CFT correspondence and the Quark-Gluon plasma,''}
  Nucl.\ Phys.\ Proc.\ Suppl.\  {\bf 216}, 94-120 (2011).
  \hri{1102.0725}{[hep-th]}.

  %\cite{Janik:2010we}
\bibitem{Janik:2010we}
  R.~A.~Janik,
  {\em ``The dynamics of quark-gluon plasma and AdS/CFT,''}
  Lect.\ Notes Phys.\  {\bf 828}, 147-181 (2011).
  \hri{1003.3291}{[hep-th]}.

%%%%%%%%%%%%%%%%%%%%%%%%%%%%%%

\bibitem{ihqcd}
 U.~Gursoy and E.~Kiritsis,
 {\em ``Exploring improved holographic theories for QCD: Part I,''}
 JHEP {\bf 0802} (2008) 032
 \hri{0707.1324}{[hep-th]};\\
 %%CITATION = JHEPA,0802,032;%%
 U.~Gursoy, E.~Kiritsis and F.~Nitti,
 {\em ``Exploring improved holographic theories for QCD: Part II,''}
 JHEP {\bf 0802} (2008) 019
 \hri{0707.1349}{[hep-th]};\\
 %%CITATION = JHEPA,0802,019;%%
 %%CITATION = FPYKA,57,396;%%

  \bibitem{gubser}
  S.~S.~Gubser, A.~Nellore,
  {\em ``Mimicking the QCD equation of state with a dual black hole,''}
  Phys.\ Rev.\  {\bf D78 } (2008)  086007.
  \hri{0804.0434}{[hep-th]}.

\bibitem{gkmn}
U.~Gursoy, E.~Kiritsis, L.~Mazzanti and F.~Nitti,
 {\em ``Deconfinement and Gluon Plasma Dynamics in Improved Holographic QCD,''}
 Phys.\ Rev.\ Lett.\ {\bf 101}, 181601 (2008)
 \hri{0804.0899}{[hep-th]};
 %%CITATION = PRLTA,101,181601;%%
 {\em ``Holography and Thermodynamics of 5D Dilaton-gravity,''}
 JHEP {\bf 0905} (2009) 033
 \hri{0812.0792 }{[hep-th]}.
 %%CITATION = ARXIV:0812.0792;%%

  \bibitem{ihqcdrev}
 U.~Gursoy, E.~Kiritsis, L.~Mazzanti, G.~Michalogiorgakis and F.~Nitti,
 {\em ``Improved Holographic QCD,''}
 \hri{1006.5461}{[hep-th]}.
 %%CITATION = ARXIV:1006.5461;%%



\bibitem{Gubser:2000nd}
  S.~S.~Gubser,
  {\em ``Curvature singularities: The Good, the bad, and the naked,''}
  Adv.\ Theor.\ Math.\ Phys.\  {\bf 4 } (2000)  679-745.
  \hre{hep-th}{0002160}.

%\cite{Gubser:2008pc}
\bibitem{Gubser:2008pc}
  S.~S.~Gubser, S.~S.~Pufu and A.~Yarom,
  {\em ``Entropy production in collisions of gravitational shock waves and of heavy ions,''}
  Phys.\ Rev.\  D {\bf 78}, 066014 (2008)
\hri{0805.1551}{[hep-th]}.
  %%CITATION = PHRVA,D78,066014;%%


  %\cite{Casini:2011kv}
\bibitem{Casini:2011kv}
  H.~Casini, M.~Huerta, R.~C.~Myers,
  {\em ``Towards a derivation of holographic entanglement entropy,''}
  JHEP {\bf 1105}, 036 (2011).
  \hri {1102.0440}{[hep-th]}.


  %\cite{Balasubramanian:2011ur}
\bibitem{Balasubramanian:2011ur}
  V.~Balasubramanian, A.~Bernamonti, J.~de Boer, N.~Copland, B.~Craps, E.~Keski-Vakkuri, B.~Muller, A.~Schafer {\it et al.},
  {\em ``Holographic Thermalization,''}
  Phys.\ Rev.\  {\bf D84}, 026010 (2011).
  \hri{1103.2683}{[hep-th]}.




  %\cite{Hotta:1992qy}
\bibitem{Hotta:1992qy}
  M.~Hotta and M.~Tanaka,
  {\em ``Shock wave geometry with nonvanishing cosmological constant,''}
  Class.\ Quant.\ Grav.\  {\bf 10}, 307 (1993).
  %%CITATION = CQGRD,10,307;%%

%\cite{Kovchegov:2009du}
\bibitem{Kovchegov:2009du}
  Y.~V.~Kovchegov and S.~Lin,
 {\em  ``Toward Thermalization in Heavy Ion Collisions at Strong Coupling,''}
  JHEP {\bf 1003}, 057 (2010)
 \hri{ 0911.4707}{[hep-th]}.
  %%CITATION = JHEPA,1003,057;%%

%\cite{Lin:2009pn}
\bibitem{Lin:2009pn}
  S.~Lin and E.~Shuryak,
 {\em  ``Grazing Collisions of Gravitational Shock Waves and Entropy Production in
  Heavy Ion Collision,''}
  Phys.\ Rev.\  D {\bf 79}, 124015 (2009)
 \hri{0902.1508}{ hep-th]}.
  %%CITATION = PHRVA,D79,124015;%%



  %\cite{Gubser:2009sx}
\bibitem{Gubser:2009sx}
  S.~S.~Gubser, S.~S.~Pufu, A.~Yarom,
{\em   ``Off-center collisions in AdS(5) with applications to multiplicity estimates in heavy-ion collisions,''}
  JHEP {\bf 0911}, 050 (2009).
  \hri{0902.4062}{ [hep-th]}.

 \bibitem{GR}
     I~S~Gradshteyn and I~M~ Ryzhik,
  {\em    "Table of Integrals, Series, and Products",}
     Academic Press, San Diego,
     Fifth Edition,
     1994.
     %Equations (6.645.2).


\bibitem{data}
  U.~Gursoy, E.~Kiritsis, L.~Mazzanti, F.~Nitti,
  {\em ``Improved Holographic Yang-Mills at Finite Temperature: Comparison with Data,''}
  Nucl.\ Phys.\  {\bf B820 } (2009)  148-177.
  \hri{0903.2859}{[hep-th]}.
  %%CITATION = NUPHA,B820,148;%%


  \bibitem{Eardley:2002re}
  D.~M.~Eardley and S.~B.~Giddings,
{\em   ``Classical black hole production in high-energy collisions,''}
  Phys.\ Rev.\  D {\bf 66}, 044011 (2002)
 \hre{ gr-qc}{0201034}.
  %%CITATION = PHRVA,D66,044011;%%

%\cite{Back:2002wb}
\bibitem{Back:2002wb}
  B.~B.~Back {\it et al.},
{\em   ``The Significance of the fragmentation region in ultrarelativistic heavy ion
  collisions,''}
  Phys.\ Rev.\ Lett.\  {\bf 91}, 052303 (2003)
  \hre{nucl-ex}{0210015}.
  %%CITATION = PRLTA,91,052303;%%

%\cite{Kovchegov:2010pw}
\bibitem{Kovchegov:2010pw}
  Y.~V.~Kovchegov,
  {\em ``Introduction to the Physics of Saturation,''}
  Nucl.\ Phys.\  {\bf A854}, 3-9 (2011).
 \hri{1007.5021}{ hep-ph]}.


%\cite{JalilianMarian:2005jf}
\bibitem{JalilianMarian:2005jf}
  J.~Jalilian-Marian, Y.~V.~Kovchegov,
{\em   ``Saturation physics and deuteron-Gold collisions at RHIC,''}
  Prog.\ Part.\ Nucl.\ Phys.\  {\bf 56}, 104-231 (2006).
\hre{ hep-ph}{0505052}.

%\cite{Albacete:2007yr}
\bibitem{Albacete:2007yr}
  J.~L.~Albacete and Y.~V.~Kovchegov,
  {\em ``Solving High Energy Evolution Equation Including Running Coupling
  Corrections,''}
  Phys.\ Rev.\  D {\bf 75}, 125021 (2007)
 \hri{0704.0612 }{[hep-ph]}.
  %%CITATION = PHRVA,D75,125021;%%

%\cite{Mueller:2002zm}
\bibitem{Mueller:2002zm}
  A.~H.~Mueller, D.~N.~Triantafyllopoulos,
  {\em ``The Energy dependence of the saturation momentum,''}
  Nucl.\ Phys.\  {\bf B640}, 331-350 (2002).
  \hre{hep-ph}{0205167}.

%\cite{Triantafyllopoulos:2002nz}
\bibitem{Triantafyllopoulos:2002nz}
  D.~N.~Triantafyllopoulos,
  {\em ``The Energy dependence of the saturation momentum from RG improved BFKL evolution,''}
  Nucl.\ Phys.\  {\bf B648}, 293-316 (2003).
\hre{hep-ph}{0209121}.

%\cite{Lappi:2011gu}
\bibitem{Lappi:2011gu}
  T.~Lappi,
 {\em  ``Energy dependence of the saturation scale and the charged multiplicity in pp and AA collisions,''}
  Eur.\ Phys.\ J.\  {\bf C71}, 1699 (2011).
\hri{1104.3725}{ hep-ph]}.

%\cite{Albacete:2009fh}
\bibitem{Albacete:2009fh}
  J.~L.~Albacete, N.~Armesto, J.~G.~Milhano and C.~A.~Salgado,
{\em   ``Non-linear QCD meets data: A global analysis of lepton-proton scattering
  with running coupling BK evolution,''}
  Phys.\ Rev.\  D {\bf 80} (2009) 034031
  \hri{0902.1112}{[hep-ph]}.
  %%CITATION = PHRVA,D80,034031;%%

%\cite{Levin:2011hr}
\bibitem{Levin:2011hr}
  E.~Levin, A.~H.~Rezaeian,
	{\em   ``Gluon saturation and energy dependence of hadron multiplicity in pp and AA collisions at the LHC,''}
  Phys.\ Rev.\  {\bf D83}, 114001 (2011).
 \hri{1102.2385}{ [hep-ph]}.

%\cite{Gelis:2010nm}
\bibitem{Gelis:2010nm}
  F.~Gelis, E.~Iancu, J.~Jalilian-Marian, R.~Venugopalan,
	{\em   ``The Color Glass Condensate,''}
  Ann.\ Rev.\ Nucl.\ Part.\ Sci.\  {\bf 60}, 463-489 (2010).
 \hri{1002.0333}{ [hep-ph]}.

%\cite{Lublinsky:2011cw}
\bibitem{Lublinsky:2011cw}
  M.~Lublinsky, E.~Shuryak,
 {\em  ``Universal hydrodynamics and charged hadron multiplicity at the LHC,''}
  \hri{1108.3972 }{[hep-ph]}.

%\cite{Kowalski:2008sa}
\bibitem{Kowalski:2008sa}
  H.~Kowalski, T.~Lappi, C.~Marquet, R.~Venugopalan,
  {\em ``Nuclear enhancement and suppression of diffractive structure functions at high energies,''}
  Phys.\ Rev.\  {\bf C78}, 045201 (2008).
 \hri{0805.4071}{ [hep-ph]}.


%\cite{Dumitru:2001jn}
\bibitem{Dumitru:2001jn}
  A.~Dumitru, J.~Jalilian-Marian,
  {\em ``Scattering of gluons from the color glass condensate,''}
  Phys.\ Lett.\  {\bf B547}, 15-20 (2002).
  \hri{[hep-ph]}{0111357}.

%\cite{Albacete:2010bs}
\bibitem{Albacete:2010bs}
  J.~L.~Albacete, C.~Marquet,
 {\em  ``Single Inclusive Hadron Production at RHIC and the LHC from the Color Glass Condensate,''}
  Phys.\ Lett.\  {\bf B687}, 174-179 (2010).
 \hri{ 001.1378}{ [hep-ph]}.

%%%%%%%%%%%%%
%\cite{Kutak:2011rb}
\bibitem{Kutak:2011rb}
  K.~Kutak,
 {\em  ``Gluon saturation and entropy production in proton proton collisions,''}
  Phys.\ Lett.\  {\bf B705}, 217-221 (2011).
  \hri{1103.3654}{[hep-ph]}.
%%%%%%%%%%%%


  \bibitem{GK}
 B.~Gouteraux and E.~Kiritsis,
 {\em ``Generalized Holographic Quantum Criticality at Finite Density,''}
 \hri{1107.2116}{[hep-th]}.
 %%CITATION = ARXIV:1107.2116;%%

 %\cite{Ochs:1996yf}
\bibitem{Ochs:1996yf}
  S.~Ochs, U.~W.~Heinz,
{\em   ``Entropy production by resonance decays,''}
  Phys.\ Rev.\  {\bf C54}, 3199-3211 (1996).
 \hre{[hep-ph]}{9606458}.

%\cite{Muller:2011ra}
\bibitem{Muller:2011ra}
  B.~Muller, A.~Schafer,
 {\em  ``Entropy Creation in Relativistic Heavy Ion Collisions,''}
  \hri{1110.2378}{[ hep-ph]}.

  \bibitem{bachas}
  C.~Bachas,
  {\em ``On the breakdown of perturbation theory,''}
  Theor.\ Math.\ Phys.\  {\bf 95 } (1993)  491-498.
  \hre{[hep-th]}{9212033}.

%\cite{Collaboration:2011hd}
\bibitem{Collaboration:2011hd}
  A.~Collaboration,
{\em  ``Kshort and Lambda production in pp interactions at sqrt(s) = 0.9 and 7 TeV
  measured with the ATLAS detector at the LHC,''}
  \hri{1111.1297}{ [hep-ex]}.
  %%CITATION = ARXIV:1111.1297;%%

%\cite{Khachatryan:2010us}
\bibitem{Khachatryan:2010us}
  V.~Khachatryan {\it et al.} [ CMS Collaboration ],
{\em   ``Transverse-momentum and pseudorapidity distributions of charged hadrons in pp collisions at sqrt(s) = 7 TeV,''}
  Phys.\ Rev.\ Lett.\  {\bf 105}, 022002 (2010).
  \hri{1005.3299 }{[hep-ex]}.


  %\cite{Khachatryan:2010nk}
\bibitem{Khachatryan:2010nk}
  V.~Khachatryan {\it et al.} [ CMS Collaboration ],
 {\em  ``Charged particle multiplicities in pp interactions at sqrt(s) = 0.9, 2.36, and 7 TeV,''}
  JHEP {\bf 1101}, 079 (2011).
  \hri{1011.5531}{ [hep-ex]}.
  %%CITATION = JHEPA,1101,079;%%

  %\cite{Collaboration:2011rta}
\bibitem{Collaboration:2011rta}
  A.~T.~f.~Collaboration,
 {\em  ``Bulk Properties of Pb-Pb collisions at sqrt(sNN) = 2.76 TeV measured by
  ALICE,''}
  \hri{1107.1973} {[nucl-ex]}.
  %%CITATION = ARXIV:1107.1973;%%


%\cite{Steinberg:2004vy}
\bibitem{Steinberg:2004vy}
  P.~Steinberg,
{\em  ``Landau hydrodynamics and RHIC phenomena,''}
  Acta Phys.\ Hung.\  A {\bf 24} (2005) 51
\hre{nucl-ex}{0405022}.

%\cite{Kiritsis:2009hu}
\bibitem{disec}
  E.~Kiritsis,
  {\em ``Dissecting the string theory dual of QCD,''}
  Fortsch.\ Phys.\  {\bf 57 } (2009)  396-417.
  \hri{0901.1772}{[hep-th]}.
  %%CITATION = FPYKA,57,396;%%

%\cite{daRocha:2006ei}
\bibitem{daRocha:2006ei}
  R.~da Rocha, C.~H.~Coimbra-Araujo,
{\em ``Extra dimensions in LHC via mini-black holes: Effective Kerr-Newman brane-world effects,''}
  Phys.\ Rev.\  {\bf D74}, 055006 (2006).
\hre{[hep-ph}{0607027]}.

\end{thebibliography}
\end{document}